\documentclass[floats,12pt]{article}
\usepackage[numbers,sort&compress]{natbib}
\usepackage{authblk}
\usepackage{amssymb,amsmath}
\usepackage[margin=1in]{geometry}
\usepackage{mathptmx}
\usepackage[english]{babel}
\usepackage{graphicx}
\usepackage{floatrow}
\usepackage{color}
\usepackage{setspace}

\begin{document}

\title{Markov Chain Models of Refugee Migration Data}

\author[1]{Vincent Huang}
\affil[1]{\small Plano West Senior  High School, Plano, Texas, 75093, USA.}
\author[2,3]{James Unwin\footnote{ Email: {\em unwin@uic.edu}}}
\affil[2]{\small University of Illinois at Chicago, Chicago, Illinois, 60607, USA.}
\affil[3]{\small Simons Center for Geometry and Physics, Stony Brook, New York, 11794, USA.}

\date{  \today}

	
		\maketitle

\abstract{The application of Markov chains to modelling refugee crises is explored, focusing on local migration of individuals at the level of cities and days. As an explicit example we apply the Markov chains migration model developed here to UNHCR data on the Burundi refugee crisis. We compare our method to a state-of-the-art ``agent-based'' model of Burundi refugee movements, and highlight that Markov chain approaches presented here  can improve the match to data  while simultaneously being more algorithmically efficient.}


\vspace{2mm}
\section{Introduction}

A record 65.6 million refugees were forcibly displaced in 2017 \cite{UNHCR}. The economic disparities between different parts of the world, and spontaneous violent ethnic or political upheavals, have made migration one of the most significant issues of the modern age.  
Notably, the civil war in Syria triggered large numbers of refugees destined for Europe or neighboring Arab nations; the ethnic cleansing of Rohingya Muslims in Myanmar led to many Rohingyas fleeing to adjacent nations such as Bangladesh; and armed conflict in Somalia, coupled with widespread drought and famine, scattered refugees throughout other nations in the Horn of Africa. 

Accurate models of refugee movements would in principle allow us to  predict the number of refugees who will arrive at a particular area or city, the date at which they will arrive, and the distribution of these refugees across multiple regions, a few days or even weeks in advance. This would allow refugee and governmental organizations to determine where to best distribute aid resources to maximize impact and efficiency.

\newpage

Migration has been studied since the 19th Century along the lines of general heuristics and simple ``gravity'' models  \cite{Ravenstein,Lee,Willekens}, inspired by Newton's theory of gravity.  More recently, a variety of algorithmic approaches have been proposed, e.g.~\cite{Christou,Groen,Klabunde,Gulden,Suleimenova,Kniveton,Johnson}, many based on agent-based Model approaches \cite{Macal} which bring significant improvements in their predictive capabilities. In this work we outline an alternative approach to simulating modern refugee crises by modeling the migration of refugees via stochastic matrices or Markov processes (see e.g.~\cite{Gagniuc,Isaacson}). The models developed here more accurately predicts the local movements of refugees in near-real time and improves on existing models in the literature. To this end we have developed for modern refugee crises on a regional level in near-real time, applied it to real world data and demonstrated that it runs efficiently from an algorithmic standpoint. 

This type of Markov chain modelling has been previously successfully applied to a variety of topics, a small selection include viral epidemics \cite{Baumann},  the internet \cite{internet} (notably, Google's PageRank algorithm \cite{PageRank} which determines the relevance of webpages),  financial systems \cite{finance}, and evolutionary biology  \cite{Djordjevic}. Markov chain models have also been applied to study large-scale problems of long-distance and more global migration patterns \cite{Blumen,Salkin,Goodman,Brown,Hirst,Nagurney,Pan,kim}, but to our knowledge this presents the first application to local movements of people and in particular to the study of refugees. We shall argue that Markov chains are an efficient and powerful manner to model refugee  movements and present an alternative to popular agent-based models. 

This work is structured as follows: In Section \ref{sec2}, we outline the geographical component of our approach to modelling refugee movements. In Section \ref{sec3}, we outline the algorithmic component of the Markov chain migration model. In Sections \ref{sec4} \& \ref{sec5}, we apply several versions of the MC Migration Model to the recent refugee crisis in Burundi and present results. In Section \ref{sec6}, we compare our simulations with results from an existing model of the Burundi refugee crisis. In Section \ref{sec7} we provide a summary and some  concluding remarks.


\section{Graphs from Geography}
\label{sec2}
The first key component in modelling refugee movements is accurately encapsulating geographical information from the region of interest. The most common approach is to construct a weighted graph $G$, where the vertices $\{v_1,v_2,\dots, v_n\}\in G$ represent cities  in the region and the edges in the graph represent roads between cities, weighted by the lengths of the roads. However, this leaves some degree of freedom in choosing which roads and cities to include, and how to assign weights; moreover refugee crises often occur in underdeveloped nations where population centers are small and the geographic data incomplete. 
For instance, one approach to determining which urban centers are significant enough to be included as vertices in the graph when reliable population data is unavailable is to utilizes nighttime images from satellites along with population estimates \cite{Gulden}.
 
A geographic map is a projection of some physical region to $\mathbb{R}^2$ and cities can be represented as points in $\{v_1,v_2,\dots, v_n\}\in \mathbb{R}^2$. Subsequently, we construct a planar graph $G_P$ by identifying the vertices as $\{v_1,v_2,\dots, v_n\}$ and introducing edges which correspond to the most major roads in the region in such a manner that any two edges intersect at their endpoints only. As a result, edges typically only link cities with other cities in their immediate vicinity. 

\newpage

An edge connecting two cities $v_i$ and $v_j$ is assigned a weight $w_{ij}$ corresponding to the physical distance in kilometers between  $v_i$ and $v_j$, as determined via Google Maps. The graph construction thus far follows that of a ``Local Interaction Model'' detailed in \cite{Gulden}, however, since many refugees will move along minor roads not included in this most basic planar graph, we will refine this initial graph construction with certain heuristics detailed below.


For any two cities $v_i,v_j$, let $d(i,j)$ be the minimal distance between $v_i,v_j$ using edges from $G_P$, meaning $d(i,j)$ is the length of the shortest path from $v_i$ to $v_j$. 
 We propose that there is some characteristic maximum distance $D$ that a refugee can travel in one day. The value of $D$ must be assigned with intuitive reasoning, rather than some prescribed method, and should incorporate information such as the geography of the region and the quality of roads. The value of $D$ is significant because it allows one to distinguish between different refugee movements based on their distance. 
We construct a refined graph $G$ of a given geographical region with cities $\{v_1,v_2,\dots, v_n\}$ by adding additional edges to $G_P$. Specifically, for each non-adjacent pair $v_i,v_j\in G_P$ with $D\le d(i,j)\le 2D$, an additional edge $\{v_i,v_j\}$ with weight $w_{ij}=D$ is introduced in the planar graph $G_P$. The resulting construction $G$ is typically non-planar.

The Floyd-Warshall algorithm \cite{FW1,FW2,FW3} can be utilized to efficiently construct this new graph. The Floyd-Warshall algorithm takes a weighted graph with vertices $v_1,v_2,\dots, v_n$ such that each pair of vertices $v_i,v_j$ is connected by an edge $e_{ij}$ of weight $w_{ij}$; if $v_i,v_j$ are not adjacent, we adopt the convention $w_{ij}=\infty$. The algorithm then computes the shortest paths between every pair of vertices $v_i,v_j$. It accomplishes this by letting $f(i,j,k)$ be the minimum length of a path between vertices $v_i,v_j$ such that any intermediate vertices $v_l$ satisfy $l\le k$. Then $f(i,j,0)$ is $w_{ij}$ because there can be no intermediate vertices and one can recursively calculate
\begin{equation}
f(i,j,k) = \min (f(i,j,k-1), f(i,k,k-1) + f(j,k,k-1)),
\end{equation}
 because the first term is the shortest valid path from $v_i$ to $v_j$ which does not pass through $v_k$, while the second term is the shortest valid path from from $v_i$ to $v_k$ and then from $v_k$ to $v_j$. The recursive step takes $O(1)$ time to compute each new value $f(i,j,k)$ for $1\le i,j,k\le n$ in terms of previously stored values $f(i,k,k-1),f(j,k,k-1)$, so the entire algorithm runs in $O(n^3)$ time. Then $d(i,j)$ is simply $f(i,j,n)$, so all minimal distances are computed.  To our knowledge, this approach of distinguishing between different pairs of vertices using their minimal distance in a planar graph has not been applied to migration before.


\section{Markov Processes and Refugee Movements}
\label{sec3}

In what follows we will combine the graph constructions outlined in the previous section with novel modelling approaches which we discuss next in order to generate a new class of migration models, whose application we explore in Section \ref{sec4}.
Consider a graph  $G$ with vertices $v_1,v_2,\dots, v_n$ such that some pairs of vertices $v_i,v_j$ are connected by edges $e_{ij}$ of weight $w_{ij}$. We demonstrated our approach to constructing such a graph which represents a geographical region in Section \ref{sec2} and we next explain how migration can be simulated on such a graph. To do this we adapt some ideas and conventions developed in previous approaches to modelling refugee movements using agent-based models, due to Groen \cite{Groen}.

The vertex in the graph indicate specific geographical locations. To model refugee migration it is useful to differentiate these nodes between three types of location, specifically
\begin{itemize}
\item Camps: Refugees in a refugee camp have a high probability of remaining in the camp, and refugees near a refugee camp have a high probability of migrating into that camp.
\item Conflict: Refugees in a conflict site have a high probability of leaving, and refugees near a conflict site have a low probability of entering the conflict site.
\item Neutral: A neutral vertex is a location which is neither a refugee camp or conflict site. Refugees in a neutral city have a moderate probability of entering or leaving.
\end{itemize}
A refugee camp is a location set up to house refugees and a conflict site is a city which has been the location of an armed conflict or other disaster. In principle a location can change type during the course of the simulation, for instance, as new refugee camps are established.
Each vertex in our model carries this `type' information as a label, along with three other labels:
\begin{itemize}
\item[{i).}] Location name;
\item[{ii).}] location type (Camp/Conflict/Neutral);
\item[{iii).}] Time stamp $t\in\mathbb{N}$;
\item[{iv).}] Refugee population at a given time stamp $N(t)\in\mathbb{N}$.
\end{itemize}
To simulate the movement of refugees the population value $N(t)$ of each vertex evolves between each time stamp. Our  model iterates through a series of discrete timestamps $t=0,1,2,\dots$ with each timestamp representing a single day and $t=0$ corresponding to the initial configuration of the system. The premise of our model is that on each day, refugees make a series of moves between locations in a region; once the total distance a refugee has travelled in a single step reaches or exceeds the maximum threshold $D$, the refugee cannot make any more movements on that day.  Between timestamps the population of each vertex changes due to the movements of the refugees.

The path that a refugee takes through the system is determined by the maximum threshold $D$ and also the properties of the vertices. Specifically, in our model a refugee at vertex $v_i$ has a probability of $R_i$ of remaining in that state. We assume that $R_i$ depends only on status of $v_i$ as either a camp, conflict, or neutral site.  In addition to this we next define the probabilities that a refugee at a given location migrates to a different location between timestamps.  
Consider two consecutive timestamps $t'$ and $t'+1$ and let $v_i$ and $v_j$ represent two adjacent vertices in $G$.
For a refugee at vertex $v_i$ at timestamp $t'$ which is able to migrate we denote by $I_{ji}(t')$ the probability that a refugee at $v_i$ for $t=t'$ moves to $v_j$ at $t=t'+1$. We refer to $I_{ji}(t)$ as the {\em intermediate probability} at timestamp $t$.

Note that $I_{ii}(t)$ represents the probability that the refugee remains in city $v_i$, so $I_{ii}(t) = R_i$.  Furthermore, we assume that a negligible number of refugees leave the system of urban centers, so the laws of probability imply that $I_{1i}(t) + I_{2i}(t) + \dots + I_{ni}(t) = 1$ for each $1\le i\le n$ where $n$ is the number of vertices in $G$. 
    It follows that 
    \begin{equation}
    \sum_{1\le j\le n,~i\neq j} I_{ji}(t) = 1- R_i.
\label{eq2}    \end{equation} 
In our model the probability that a refugee will move to a particular location is a function of the edge weight $w_{ij}$ (which encodes the geographical distances) and a  scaling factor $\alpha$ that takes different constant values depending on whether the location is a camp, conflict or neutral site (we specify the values of $\alpha$ in Section \ref{sec4} for a specific refugee crisis).
For each vertex $v_j$ adjacent to $v_i$ we define the  intermediate probability to be 
\begin{equation}
I_{ji}(t)=\frac{\alpha\cdot\beta}{w_{ij}},
\end{equation}
  where $\beta$ is a constant scaling factor fixed such that the probabilities add to unity after the type rescaling, thus eq.~(\ref{eq2}) is satisfied.

This setup has the following desired properties:
\begin{itemize}
    \item The total sum of the intermediate probabilities from any city $v_i$ is $1$.
    \item Refugees are more likely to migrate to refugee camps than to neutral cities. 
    \item Refugees are more likely to migrate to neutral cities than to conflict sites. 
    \item Refugees cannot directly migrate between non-adjacent cities. 
    \item The probability of moving to an adjacent city $v_j$ is inversely proportional to the distance $w_{ij}$ separating the cities. 
    \end{itemize}
Note that this last point is in accordance with Ravenstein's Laws of Migration \cite{Ravenstein}, which state that migrants prefer to move shorter distances.

Thus far we have discussed how to encode the local  preferences for refugee movements which is characterized by the intermediate probability defined above. 
We next introduce the notion of a  {\em transition probability} $P_{ji}(t)$, being the probability that a refugee in city $v_i$ at the start of timestamp $t$, will move to city $v_j$ at the start of the next timestamp $t+1$.

Our  approach maintains that a refugee may make multiple moves between cities in a single day until the refugee's total distance travelled exceeds $D$. Each individual move is made according to the intermediate probabilities $I_{ji}$, while the transition probabilities $P_{ji}$ measure the net effect of these movements taken in series. Thus the transition probabilities are completely determined by the choice of intermediate probabilities.

Suppose that at timestamp $t$ that the refugee population of city  $v_i$ is $N_i(t)$ for $1\le i\le n$, we define the $n$ dimensional `population' vector  which is a function of the timestamp $t\in\mathbb{N}$
\begin{equation}
B(t)=\{N_1(t),~N_2(t),\cdots,N_n(t)\}~.
\end{equation}
For a system with transition probabilities $P_{ij}(t)$ for $1\le i,j\le n$,  the stochastic matrix of the system at time $t$ can be denoted as
 \begin{equation}
A(t) = \begin{bmatrix} P_{11}(t) & P_{12}(t) & ... & P_{1n}(t) \\ P_{21}(t) & P_{22}(t) & ... & P_{2n}(t) \\ \vdots & \vdots & ... & \vdots \\ P_{n1}(t) & P_{n2}(t) & ... & P_{nn}(t)\end{bmatrix}~.\end{equation}
The entries of the matrix are each probabilities with the rows summing to one, thus the matrix describes transitions in a Markov chain. 

The stochastic matrix $A(t)$ dictates the probability of moving from $v_j$ to $v_i$ between timestamp $t$ and $t+1$. Thus the expected population vector at timestamp $t+1$ can be described compactly as $B(t+1) = A(t)B(t)$. Moreover, the $i$th entry of $A(t)B(t)$ gives the expected number of refugees at vertex $v_i$ at timestamp $t+1$  given by
\begin{equation}
[B(t+1)]_i = \big[A(t)\cdot B(t)\big]_i = P_{i1}(t)N_1(t) + P_{i2}(t)N_2(t) + \dots + P_{in}(t)N_n(t) ~.
\end{equation}
More generally  for a given timestamp $t'$ the vector $B(t')$ can be defined from the initial population vector $B(0)$ recursively
\begin{equation}
B(t')=A(t'-1)\cdot  A(t'-2)\cdots A(0)\cdot B(0)~.
\end{equation}
Thus at the start of each sequential timestamp the vertex populations are inherited from the final populations at the end of all operations of the previous timestamp. This is traditional Markov process of an iteratively evolving system. More specifically, since the form of $A$ is not fixed but intermittently evolves over the timestamps if a neutral city becomes a camp or conflict site, this is a non-homogeneous Markov chain \cite{Isaacson} . 

This Markov system outlined above is advantageous for modelling migration because its allows for a relatively compact model solely in terms of the transition probabilities $P_{ji}(t)$ and concerns about intermediate probabilities and about refugees moving across multiple cities in one day are irrelevant because they are already encapsulated into the transition probabilities. As a result, the main difficulty in implementing a stochastic matrix model is to compute the values of $P_{ji}(t)$. 
 Our algorithm for evolving the vertex populations between two timestamps performs  a set of order operations and thus one can break the problem of computing transition probabilities into smaller pieces. This permits the application of dynamic programming techniques \cite{Bellman} which are essentially an efficient form of recursion. 
To implement this we introduce a new location specific $n$-tuple $g_i(d,t)=\{s_1(d,t),s_2(d,t), \dots, s_n(d,t)\}_i$ associated to vertex $v_i$. The elements $s_j(d,t)$  of $g_i$ represent the probability that the refugee will be at vertex $v_j$ at the end of timestamp $t$, which depends on 
the distance $d$ a refugee has previously travelled during the current timestamp $t$ such that they currently reside in city $v_i$. 

Specifically, if $d\ge D$, then the refugee cannot make any further movements to different cities, so will remain in city $v_i$ for the remainder of the timestamp. Therefore 
\begin{equation}
g_i(d,t)\big|_{d\ge D}=\{ 0,0,\dots, 0, 1,  0, \dots, 0\}~.
\label{eqg2}
\end{equation}
 where the single $1$ occurs in the $i$th entry of the vector. 
If $d<D$, then for each $1\le j\le n$, the refugee moves from $v_i$ to $v_j$ with intermediate probability $I_{ji}(t)$; at the end of this move, the refugee is now in city $v_j$ and has moved a total distance of $d + w_{ij}$. At this point, the probability distribution for the final destination of the refugee is $g(j, d+w_{ij})$. It follows that 
\begin{equation}
g_i(d,t)\big|_{d< D} = \sum_{1\le j\le n} I_{ji}(t) g(j,d+w_{ij}).
\label{eqg}
\end{equation}
 Note that when $v_i,v_j$ are not adjacent, our sum contains a term $I_{ji}(t) g(j,d+w_{ij}) = 0\cdot g(j,\infty)$; we adopt the convention $g(j,\infty)=\{ 0,0, \dots, 0\}$ for all $j$ to ensure the expressions are well-defined, though the multiplication by zero implies the exact value of $g(j,\infty)$ is irrelevant.  

Thus eq.~(\ref{eqg}) yields a recursive formula for $g_i(d,t)$ in terms of expressions of the form $g_j(d',t)$, where $d'=d+w_{ij}>d$. As a result, we can efficiently compute all vectors $g_i(d,t)$ by first computing all vectors of the form $g_i(D-1,t)$ and storing the results, then computing all vectors of the form $g_i(D-2,t)$, and so on; eq.~(\ref{eqg2}) handles terms $g_i(d,t)$ with $d\ge D$ that occur in the recursion. Computing each $g_i(d,t)$ in terms of $n$ terms of the form $I_{ji}(t) g_j(t)(d+w_{ij},t)$ takes $O(n^2)$ time: It is essentially summing $n$ vectors of the form $g_j(d+w_{ij},t)$, each of which has already been computed and is $n$ dimensional. Therefore, this algorithm computes the $nD$ vectors of the form $g_i(d,t)$ for $1\le i\le n, d<D$ in $nD\cdot O(n^2) = O(n^3D)$ time. 

The transition probability $P_{ji}(t)$ is equivalent to the probability that a refugee currently located at city $v_i$ during timestamp $t$ who has travelled $d=0$ distance ends up at city $v_j$ by the end of timestamp $t$. As a result, $P_{ji}(t)$ is the $j$th entry in $g_i(0,t)$ for all $1\le i,j\le n$, so the computation of all $g_i(d,t)$ terms allows one to compute all the transition probabilities $P_{ji}$(t). 

The transition probabilities are constant over time unless the intermediate probabilities $I_{ji}(t)$ change which only happens when a city becomes a refugee camp or conflict site and whilst this occurs in our models, it is not a common occurrence. As a result, the above procedure should only run a few times during each refugee crisis. The computation of transition probabilities from intermediate probabilities completes the algorithmic aspect of our  model.


\section{Application: Modelling the Burundi Crisis}
\label{sec4}

\begin{figure}[t!]
       \centering
       \includegraphics[width=0.335\textwidth]{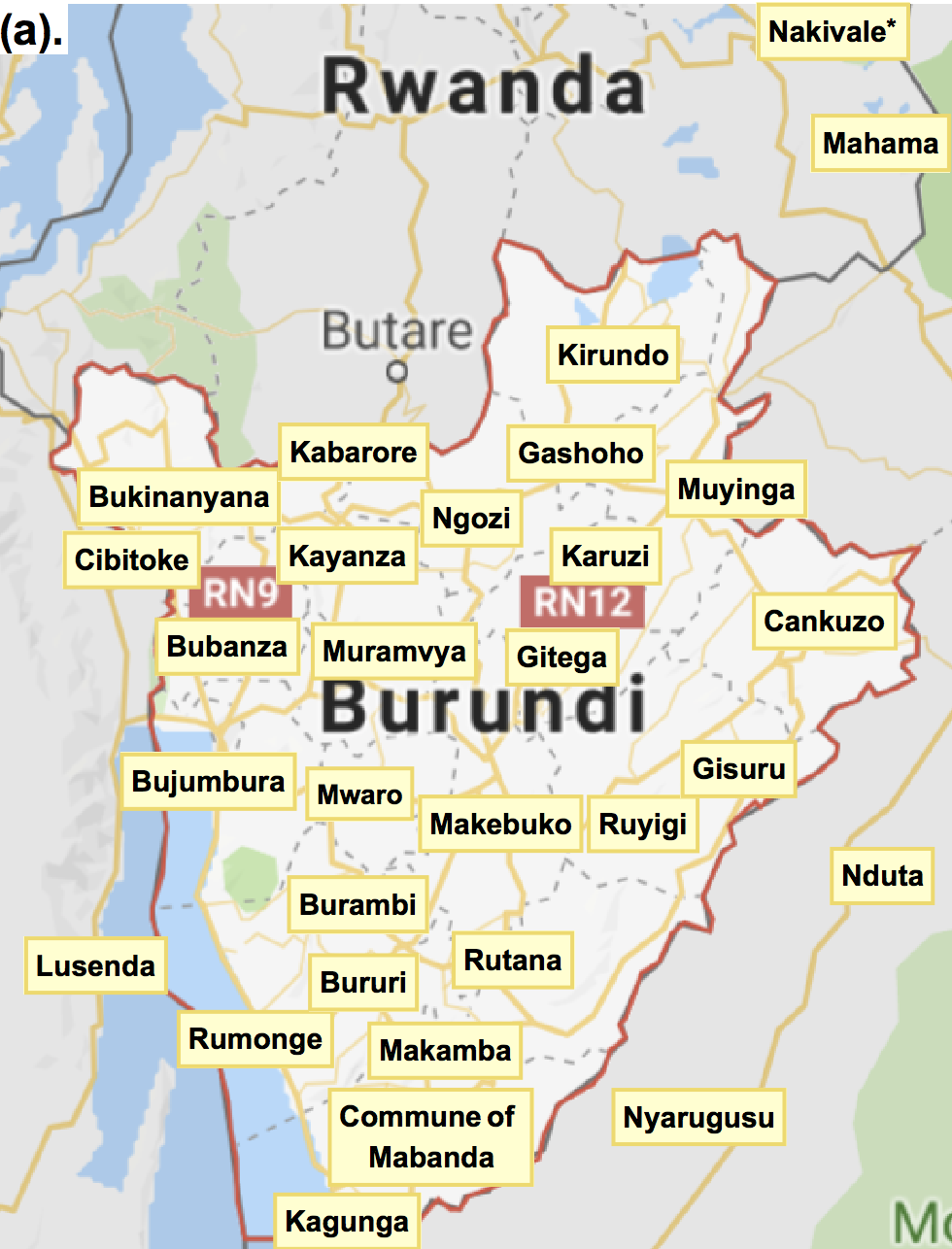}~
       \includegraphics[width=0.32\textwidth]{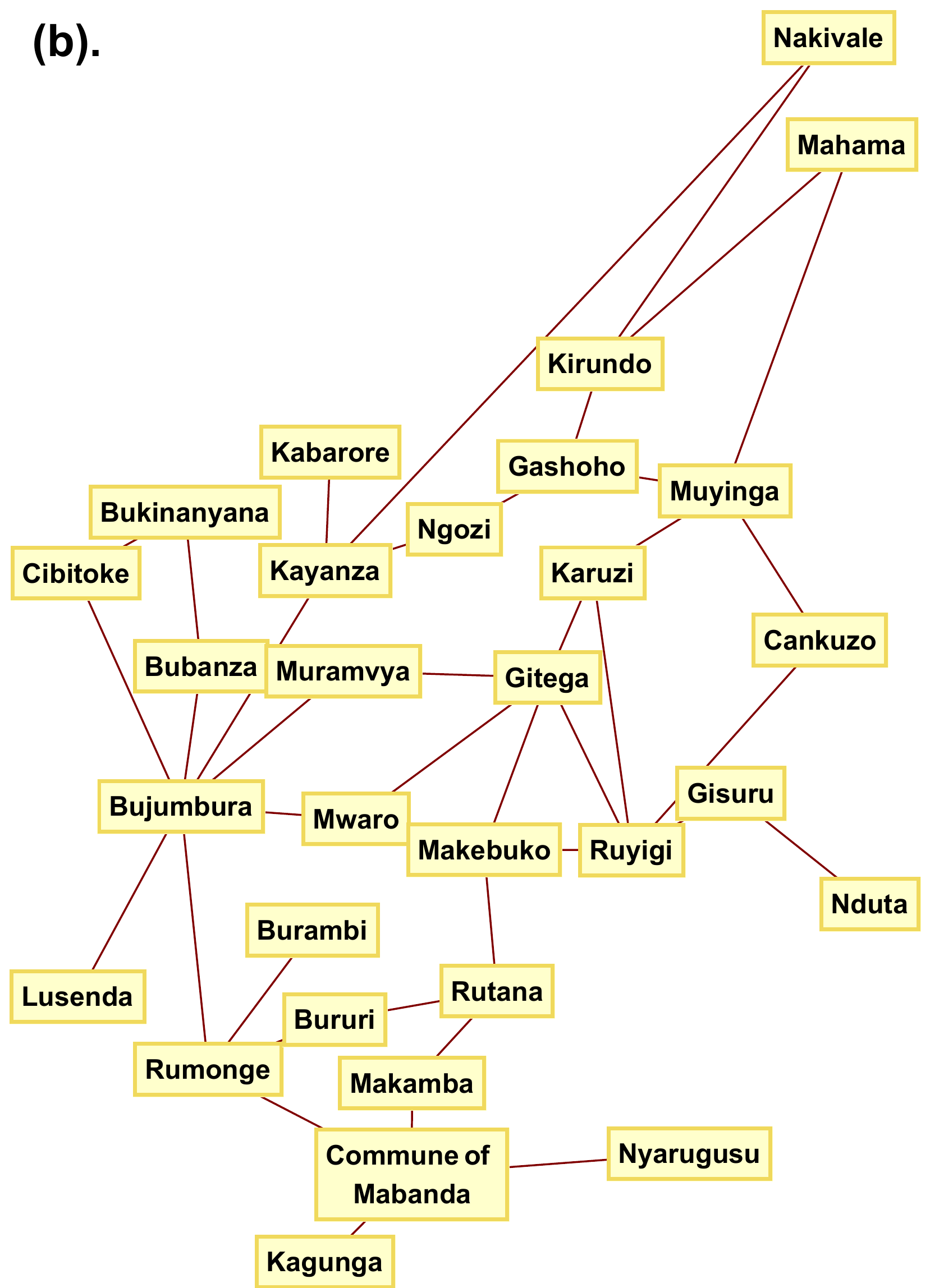}~
       \includegraphics[width=0.32\textwidth]{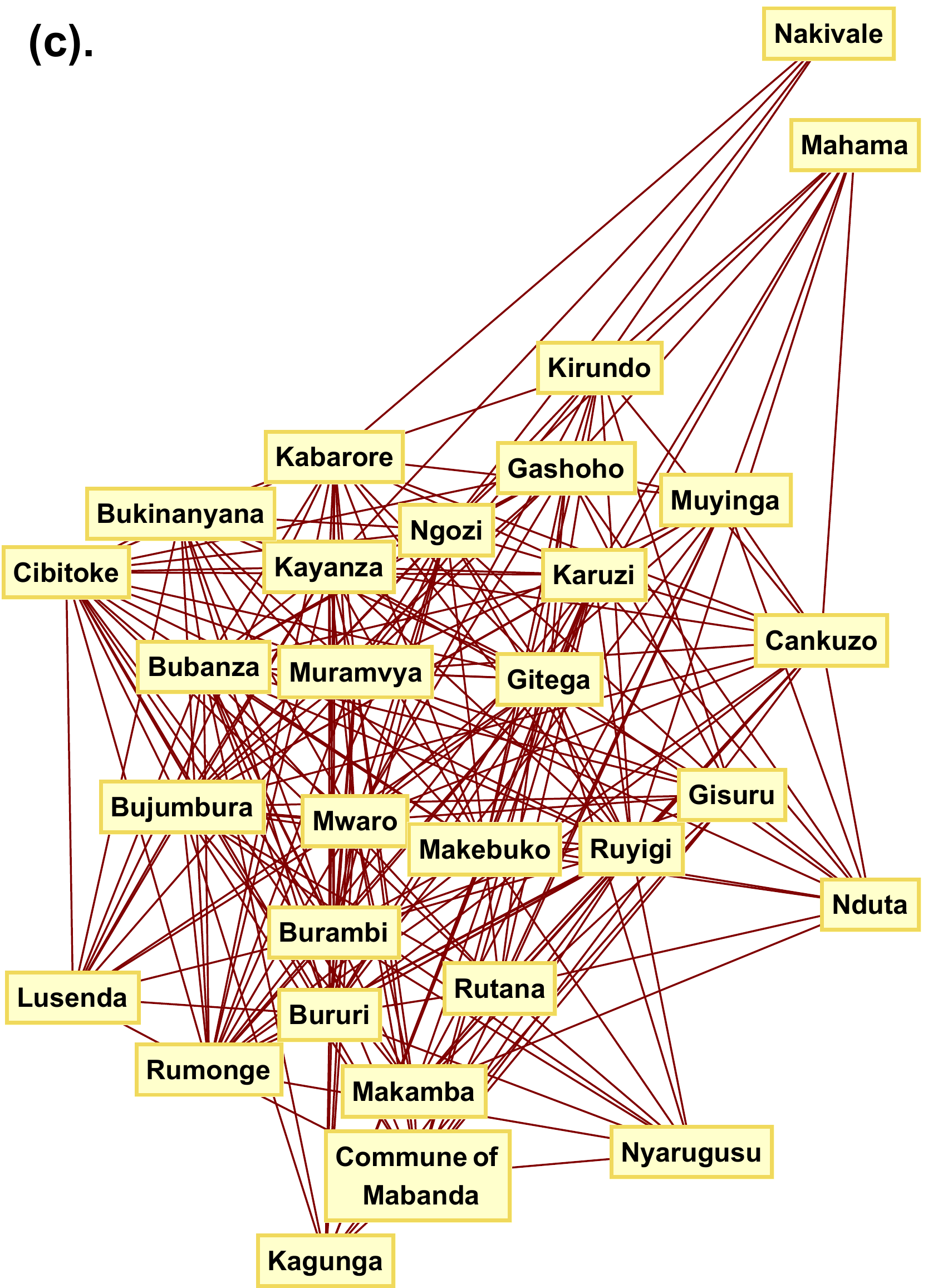}
       \vspace{-4mm}
              \caption{(a). Location labels overlaid on a map of Burundi and surrounding regions (from Google Maps \cite{google}). Note, Nakivale is located further north than illustrated. (b).~A planar graph $G_P$ representing major roads connecting locations relevant to the Burundi refugee crisis as used in  \cite{Suleimenova}. (c).~Non-planar graph representing additional routes connecting relevant locations, constructed by adding additional edges to the graph  $G_P$ in panel (b), as detailed in Section \ref{sec3}. \vspace{-3mm}} 
       \label{Fig1}
\end{figure}

 We utilize data on the Burundi crisis from a previous model of the crisis called \textit{Flee} due to Suleimenova-Bell-Groen (SBG) \cite{Suleimenova}, who in turn obtained their data from the United Nations High Commissioner for Refugees (UNHCR). The UNHCR data extracted by SBG \cite{Suleimenova} runs from May of 2015 to June of 2016, for a total of $396$ days, so the timestamps in the model range from $0\le t\le 395$. 
There were five active refugee camps in the Burundi crisis, each established by the UN at the cities of Nyarugusu, Nduta, Nakivale, Mahama, and Lusenda.  Figure \ref{Fig1}a shows the 30 major cities in and around Burundi which were modelled in Flee and Figure \ref{Fig1}b illustrates the graph used to model the system constructed and used in SBG \cite{Suleimenova}. Figure  \ref{Fig1}c shows a non-planar graph representing additional routes connecting relevant locations, which we constructed by adding further edges to the graph  $G_P$ in panel (b), as detailed in Section \ref{sec3}.

In SBG \cite{Suleimenova} the authors assumed that the probabilities $R_i$ of remaining in a city $v_i$ are 
\begin{equation}
R_i=\left\{\begin{array}{cc}
0 & v_i~{\rm is~a~conflict~site} \\
0.7& v_i~{\rm  is~a~neutral~site} \\
0.999& v_i~{\rm  is~a~camp~site}
\end{array}\right.~.
\end{equation}
These estimates seem reasonable so we maintain them in our model too. We also take a list of major armed conflicts in Burundi from SBG's {\em Flee} model \cite{Suleimenova} and turn the corresponding locations into conflict sites at the appropriate timestamp. The first such conflict took place on May 1st, 2015, which we set as the beginning date $t=0$ of the model.  SBG \cite{Suleimenova} estimate that  refugees travel a maximum distance of $200$ km in one day.\footnote{The supplementary materials of  \cite{Suleimenova} presents sensitivity testing for Flee and shows that the results are robust under modest variations of the maximum distance $D$ and the staying probabilities $R_i$.} Arguably, this overly optimistic as most refugees travel on foot \cite{Suleimenova}, so we instead set a daily distance threshold of $D=120$ (in units of km), corresponding to $15\text{ hours}$ at $\text{8 km/h}$.

Next, we initialize the population vectors $B(t)$ to match the number of refugees in the Burundi region; again we obtained this data from \cite{Suleimenova}, who retrieve it from UNHCR refugee camp registration data. SBG \cite{Suleimenova} notes that it is incredibly difficult to predict the number of refugees created by armed conflicts, but rough estimates can be obtained through various means such as satellite images \cite{Gulden}. As a result, rather than predicting the number of refugees resulting from a crisis, SBG \cite{Suleimenova} aim to predict the distribution of these refugees given the total number of them; our model also focuses on this objective. We will return in Section \ref{sec5} to directly compare our results with those from the {\em Flee} model of SBG \cite{Suleimenova}.

Since  $B(t+1)=A(t)B(t)$ represents the expected distribution at time $t+1$ of the refugees from time $t$, the total number of refugees in $B(t+1)$ is the same as the total number of refugees in $B(t)$.
This implies that any increases in the number of refugees are not reflected in the matrix multiplication and therefore must be performed through an additional process. Specifically, we denote by $U(t)$  the number of refugees in Burundi refugee camps at time $t$ according to the UNHCR. After each timestamp $t$, the population discrepancy $U(t+1)-U(t)$, between timestamp $t$ and timestamp $t+1$, is compute and then an additional $U(t+1)-U(t)$ refugees is randomly added into the population vector $B(t)$ by distributing them into the conflict sites in the model with probability proportional to the population of the conflict site. This ensures the model always contains an accurate number of refugees. 

 Due to this mechanism of adding refugees, after timestamp $t$, the total number of refugees in the model is the same as the total number of refugees in UNHCR camps after day $t$. As a result the model continually under populates the graph and therefore under predicts the number of refugees in each camp, because many refugees in the model are still making their way to camps. 
To account for this delay between the creation of refugees and their arrival in camps in the model, when the refugee populations in each refugee camp are extracted, following \cite{Suleimenova} we rescale each such population upwards by a constant factor so that the total number of refugees in the camps matches the UNHCR data for each day. This rescaling occurs in three of the four Markov chain models we outline below and in Flee \cite{Suleimenova}. Because our stated goal is to predict the distribution of refugees in  camps given the total number of refugees, this rescaling is unimportant.  However, since this scaling is somewhat ad hoc one of the Markov chain variants we outline below is designed to remove the need for this rescaling.

We now apply the modelling algorithms outlined in Sections \ref{sec2} \& \ref{sec3}  to the recent Burundi refugee crisis, due to a civil war, focusing on a set data on camp registrations during a period from 2015-2016 . We run four different models using Markov chains, each of which is a variation on the principles outlined in Sections \ref{sec2} \& \ref{sec3}, as we outline below:

\vspace{2mm}

{\bf Markov Chain: Initial graph}.
We apply the Markov Chain approach to the planar graph $G_P$ constructed in \cite{Suleimenova} from an analysis of major roads in Burundi on Bing Maps, see Figure~\ref{Fig1}b.

\vspace{2mm}

{\bf Markov Chain: Graph-adjusted}.
We additionally apply our  Markov chain approach to the modified graph $G$ which we derive from $G_P$ using the procedure from Section \ref{sec2}, show in Figure \ref{Fig1}c,  in what we call the graph-adjusted model. 

\vspace{2mm}

{\bf Markov Chain:  Camp-adjusted.}
Long movements to refugee camps are systemically assigned smaller intermediate probabilities in our  model. This heuristic makes sense for movements to neutral cities because refugees prefer making a series of shorter movements rather than a single long movement \cite{Ravenstein}, but this reasoning breaks for refugee camps because they are refugees' final destinations rather than intermediate locations. In addition, there are a variety of motivations for refugees to prefer faraway refugee camps, such as perceived safety and distance from danger. 
To realise this observation, we create an additional graph $G'$ by taking edges $e_{ij}$ in $G$ that represent long movements to refugee camps and reassigning these edges with weight $w_{ig}=D$ so that long movements to refugee camps have the same chance of occurring as medium movements to camps. We call this variation the \textit{camp-adjusted model}.

\vspace{2mm}

{\bf Markov Chain: Time-adjusted.} We now present a fourth Markov chain model, which we call the \textit{time-adjusted model}, which removes the need for the overall rescaling to match the total refugee population at the end of each timestamp. 
 Denote by $T$ the average time it takes for a refugee in the model to travel from a conflict site to a refugee camp. Then if there are $U(t+T)$ refugees in the model at timestamp $t$, most of these refugees should arrive in refugee camps within $T$ days, so the number of refugees residing in camps within the model will be approximately $U(t+T)$ at timestamp $t+T$.
We can use this observation to remove the need for the rescaling at the end of each timestamp by introducing an appropriate number of refugees in the model $T$ days before they are observed. Specifically, at timestamp $t=0$ we introduce $U(T)$ refugees instead of $U(1)$ refugees, and at the end of each timestamp $t$ we add an additional $U(t+T+1)-U(t+T)$ refugees rather than $U(t+1)-U(t)$ refugees. 

It remains to compute $T$, the average travel time for a refugee from a conflict site to a refugee camp, which can be found by running a mock simulation where the entire population $N$ of the conflict sites become refugees at timestamp $t=0$, and then let $N(\delta)$ denote the total camp population after $\delta$ days for $1\le \delta\le M$ for some suitably large constant $M$. Then $N(\delta+1)-N(i)$ refugees take exactly $\delta+1$ days to arrive, so the average travel time is
 \begin{equation*}\begin{aligned}
 T &=\frac1N \sum_{\delta=0}^{M-1} (\delta+1)(N(\delta+1)-N(\delta))\\
  &= \frac1N \Big(MN(M) -\big(N(M-1)+N(M-2)+\dots +N(0)\big)\Big).
  \end{aligned}\end{equation*}
For $M=100$ we find $T=16$ and thus this is the value we use in the time-adjusted model.

\section{Results: Modelling the Burundi Crisis}
\label{sec5}

\begin{figure}[t!]
       \centering
\vspace*{-3mm}
       \includegraphics[width=0.45\textwidth]{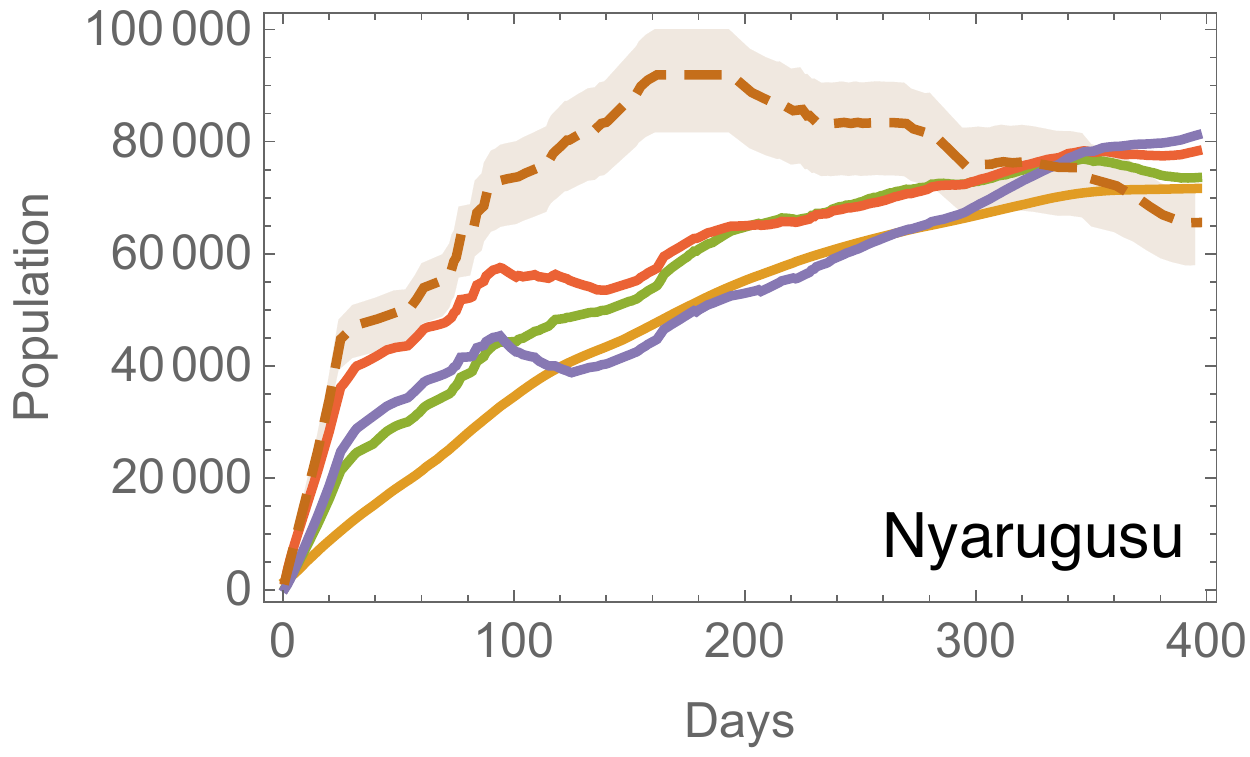}~
       \includegraphics[width=0.45\textwidth]{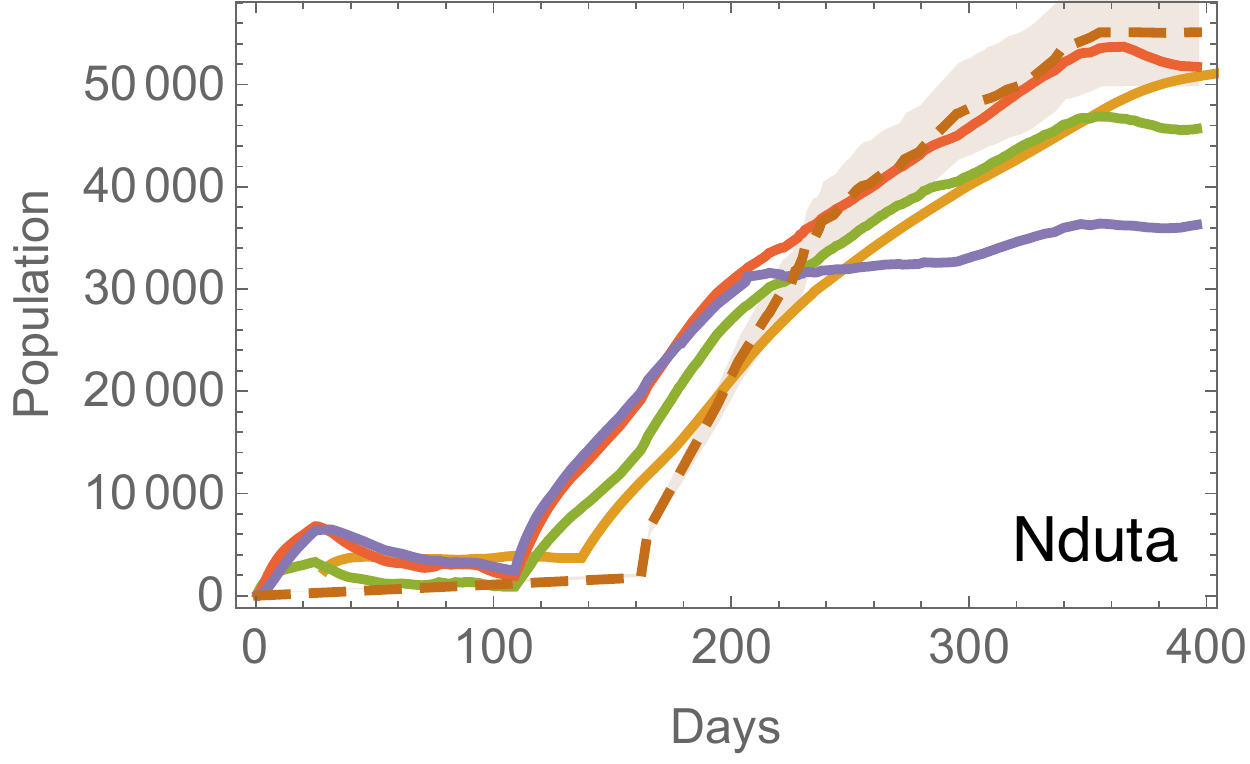}\\
       \includegraphics[width=0.45\textwidth]{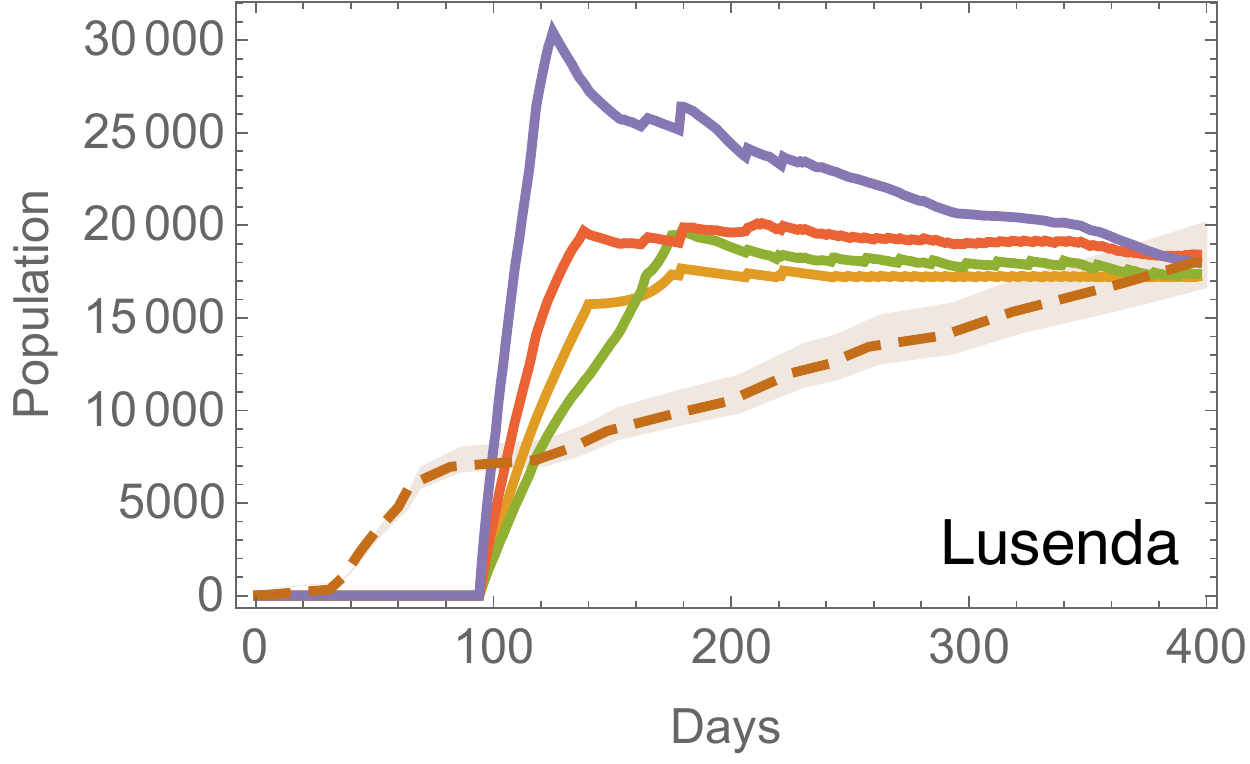}~
       \includegraphics[width=0.45\textwidth]{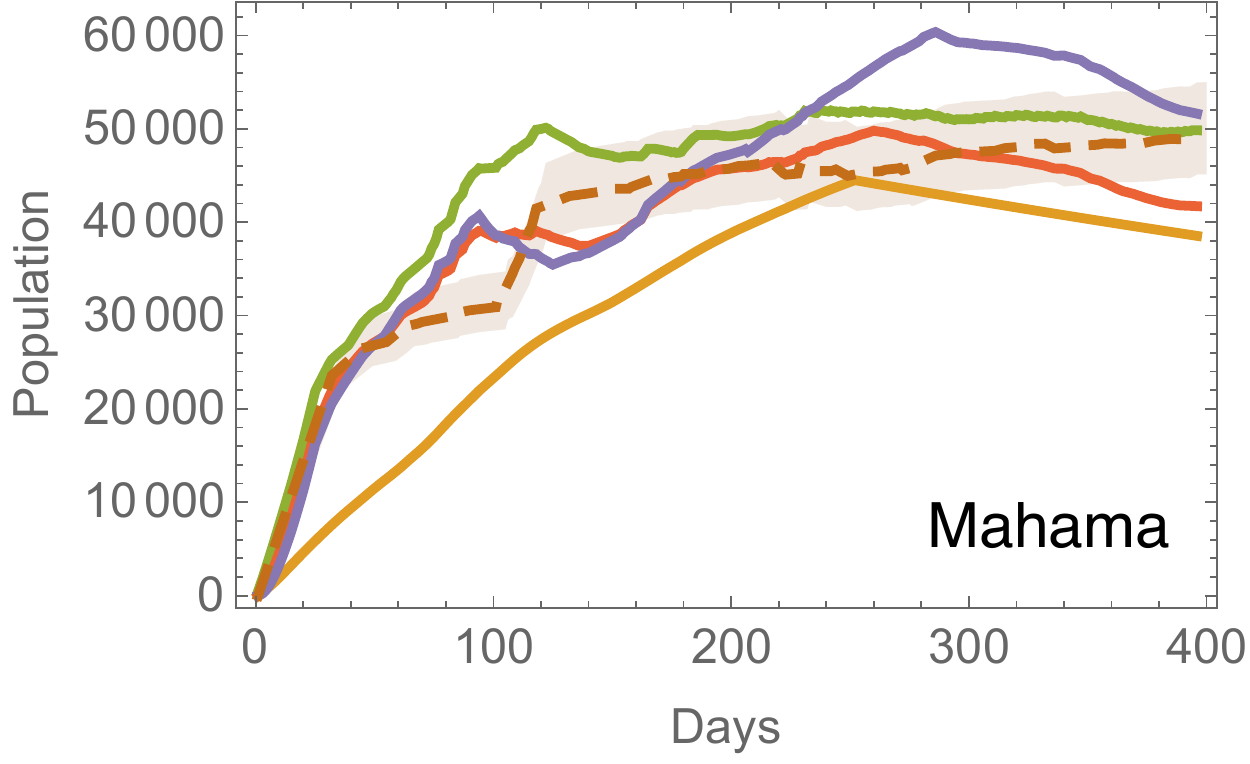}\\
       \includegraphics[width=0.45\textwidth]{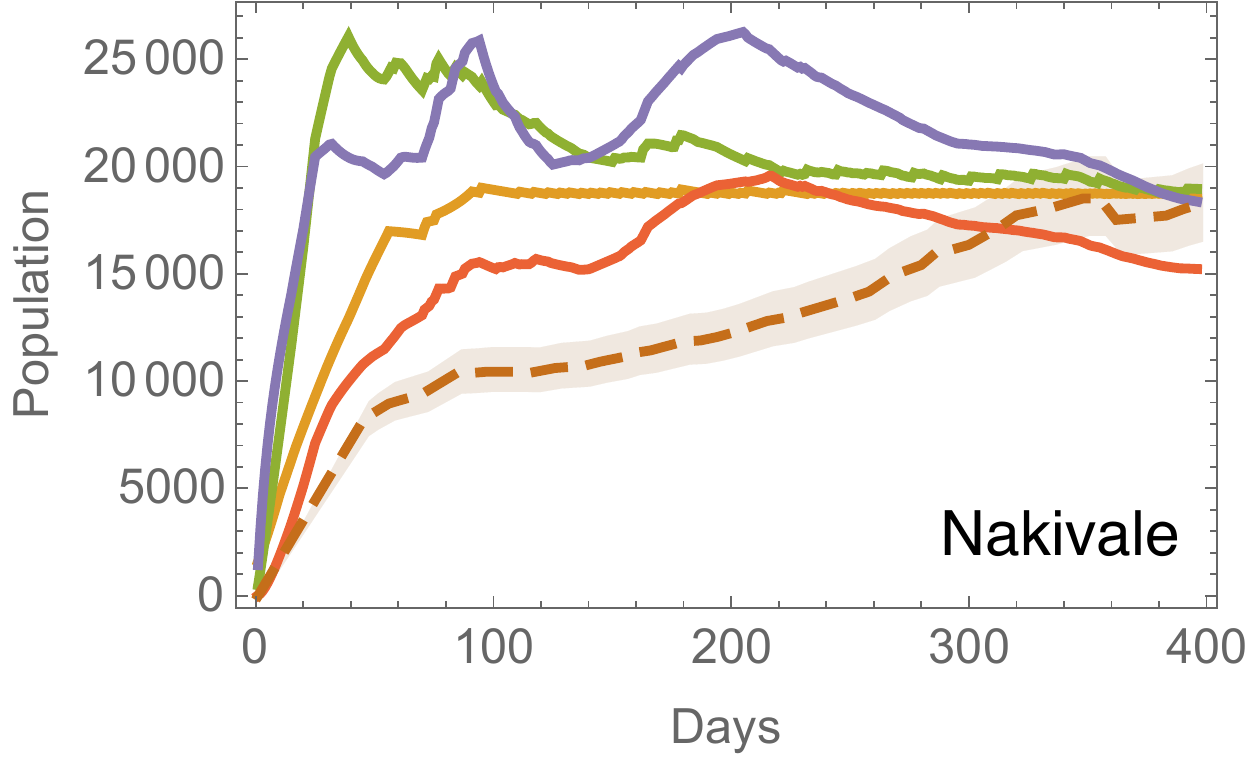}~
       \includegraphics[width=0.45\textwidth]{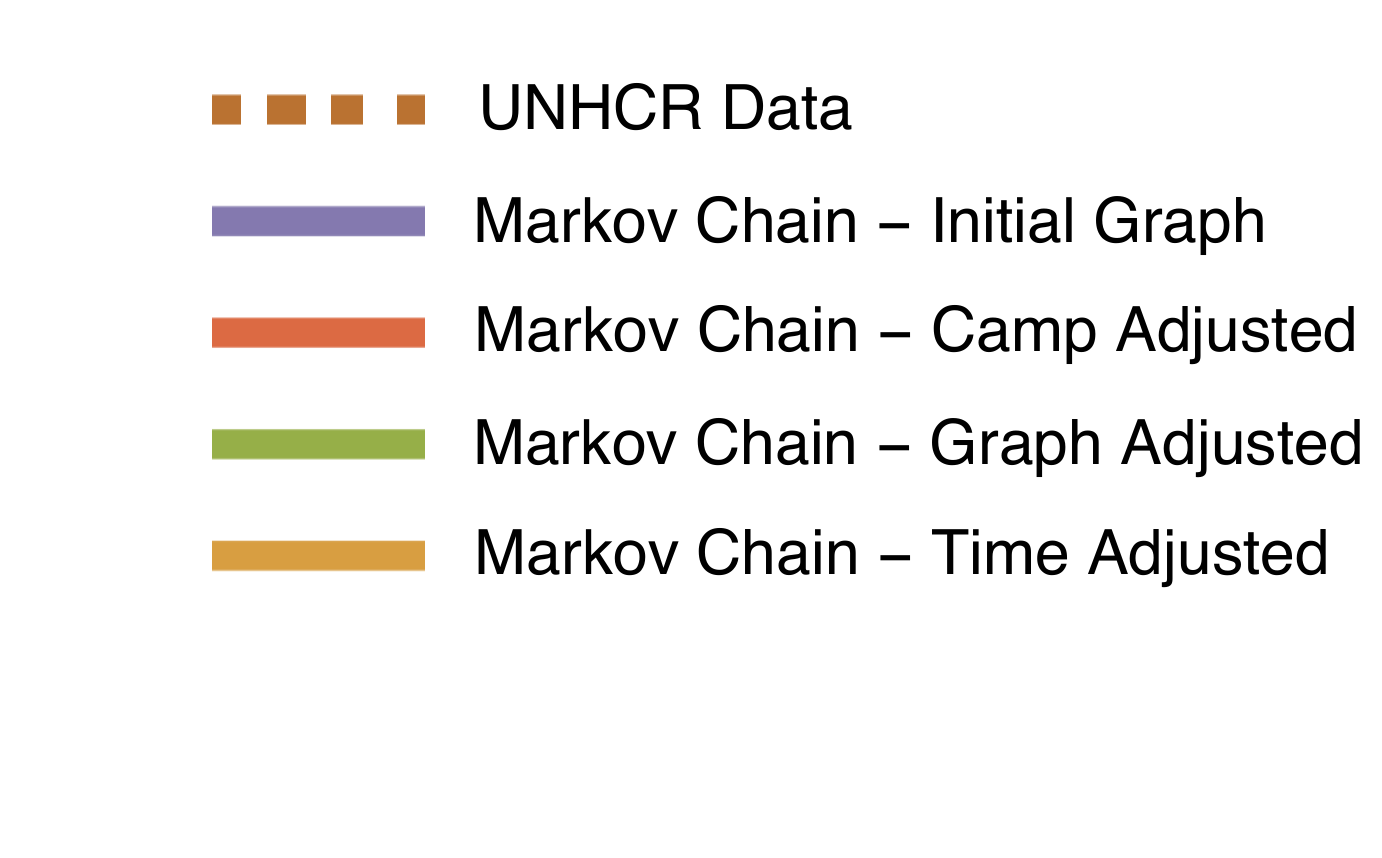}
\vspace*{-3mm}
              \caption{Plots of the population growth in each of the five major refugee camps relevant to Burundi refugee crisis predicted by each of our models as a function of timestamp (days). The four solid curves in each plot correspond to the four Markov chain models outlined in Section \ref{sec4} and the dashed curve is the UNHCR  camp data from  \cite{Suleimenova}, the shading indicates a 10\% error in the data. \vspace{-2mm}} 
       \label{Fig2}
\end{figure}

We will now present the results of the Burundi models outlined above.  Figure \ref{Fig2} shows five different graphs which are model results for number of refugees in the major regional camp as a function of time. Each graph represents one of the five  different camps relevant to the Burundi crisis, and in each graph we plot the refugee population in the given camp as a function of the timestamp (days). Within each graph we show five curves: the four solid curves correspond to the four Markov chain models outlined in Section \ref{sec4}, the dashed curve is the UNHCR camp population data, and the shaded region indicates a 10\% error on the data.

It is difficult to determine from these population plots alone how effective each model is at predicting refugee populations in the five camps. To quantify the goodness of fit we now introduce metrics to measure the success of a given refugee model.  Denote by $X$  the set $\{\text{Nyarugusu}, \text{Nduta}, \text{Nakivale}, \text{Mahama}, \text{Lusenda}\}$ of relevant refugee camps. For each camp $x_i\in X$, let $N_i(t)$ be the population of that camp at timestamp $t$ in a given model, and let $\pi_i(t)$ be the population of that camp at day $t$ according to UNHCR data. Note that $U(t)$ is precisely the sum of $\pi_i$ over each of the camps: $\sum_{x_i\in X} \pi_i(t) = U(t)$. 
The first measure we consider is the Averaged Relative Difference (ARD) for a given camp $x_i$ on day $t$, given by
\begin{equation}
E_i(t) = \frac{\vert N_i(t) - \pi_i(t) \vert}{U(t)}.
\label{ARD}
\end{equation}
The total ARD on a given timestamp $t$ is then the sum over the ARD values of all of the camps 
\begin{equation}
E(t) = \sum_{x_i \in X} E_i(t).
\label{TARD}
\end{equation}

Moreover, it is also useful to consider time averaged ARDs for a given model, by averaging over all the daily ARD values in a certain period. In particular, we consider the Average ARD  $\langle E\rangle$ over the entire period covered by the data (up to $t=396$)
\begin{equation}
\langle E\rangle =  \frac{1}{396}  \sum_{j=0}^{395} E_j=  \frac{1}{396} (E(0) +E(1) + \dots +E(395))~.
\label{395}
\end{equation}

\vspace{-5mm}
\begin{figure}[t!]
       \centering
       \includegraphics[width=0.44\textwidth]{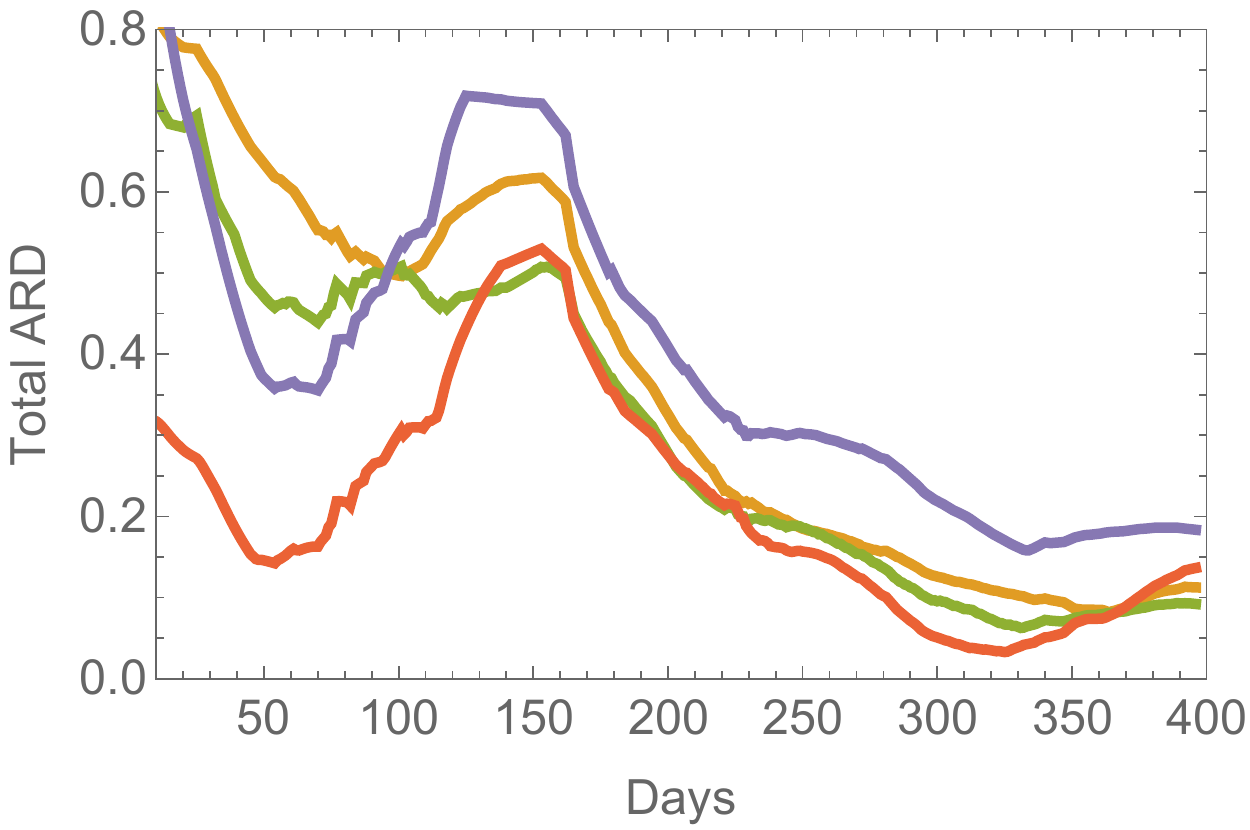}
       \caption{Total  Averaged Relative Difference (ARD), as defined in eqns.~(\ref{ARD}) \& (\ref{TARD}), on each timestamp (day) for each of the Markov chain models. It can be seen that the refined graphs offer an improved fit of the data. Coloured curves match those in Figure \ref{Fig2}. } 
       \label{Fig3}
\end{figure}

\clearpage

Figure \ref{Fig3}  shows the total ARD $E(t)$ for each model as a function of time. This provides a good comparison tool to assess how well the models match the data. Observe that the Markov chain model using the `Initial graph' (Figure \ref{Fig1}b) consistently has relatively high error margins, while the {\em graph-adjusted} and {\em camp-adjusted} models  (Figure \ref{Fig1}c) are substantially improved.

Notably, there is potentially a lot of noise in the first few days of every model and in actual refugee crises, thus we expect the quality of the data to improve over time, as registration practices improve and the situation in the camps becomes more stable. To better analyze the long-term accuracy of the models we also consider the averaged ARD  omitting the earliest period of the data, since this likely provides a more reliable measure of `goodness of fit'. Specifically, we call the Average ARD omitted the first $z$ days (counting from day $t=0$) the $z$-day Clean Averaged ARD and denote this by $\langle E\rangle_z$ with
\begin{equation}
\langle E\rangle_{z} \equiv  \frac{1}{396-z}  \sum_{j=z}^{395} E_j ~.
\end{equation}
We will consider the 30-day and 100-day Clean Averaged ARD, denoted $\langle E\rangle_{30}$  and  $\langle E\rangle_{100}$, the values of which are shown in Table \ref{tab1} for the four Markov chain models of Section \ref{sec4}.

\vspace*{3mm}
\begin{table}[H]
\begin{tabular}{|c|c|c|c|c|c|}
\hline
        & Initial Graph &  Camp-Adjusted     &  Graph-Adjusted &  Time-Adjusted \\ \hline
$\langle E \rangle$   & 0.41 & 0.22        & 0.32       & 0.37       \\ \hline
$\langle E \rangle_{30}$  & 0.38 & 0.21        & 0.28       & 0.34      \\ \hline
$\langle E \rangle_{100}$ & 0.37 & 0.21       & 0.24       & 0.28      \\ \hline
\end{tabular}
\caption{All time Average ARD $\langle E\rangle$, and Clean Average ARDs $\langle E\rangle_{30}$  and $\langle E\rangle_{100}$ for each model.}
\label{tab1}
\end{table}

From  the statistics from Table \ref{tab1} we can draw some general conclusions about our  models. Firstly, the Camp-Adjusted  Model has a lower Average ARD than every other model in each of the three numerical measures  $\langle E \rangle$, $\langle E \rangle_{30}$, and $\langle E \rangle_{100}$. Furthermore, the results clearly demonstrate the merit of the geographical adjustments introduced in Section \ref{sec2}, with the Camp-Adjusted and Graph-Adjusted  Models yielding the lowest long-term average ARDs. 
That the Camp-Adjusted and Graph-Adjusted  models give an improved match to data compared to graphs which represent the major road system may be indicative of refugees travelling off-road, or on minor roads not well documented on Bing or Google Maps. It would be interesting to further validate this conclusion over additional conflicts and longer term data sets.

 The Time-Adjusted  Model begins with relatively inaccurate results, but the average error quickly drops, as expected because this is a time-sensitive model. However, omitting the first 100 days and considering $\langle E\rangle_{100}$, the Time-Adjusted  Model gives a comparably low average ARD to the other models, indicating that we were successfully in removing the rescaling of refugee camp populations at the end of each timestamp while still producing a relatively accurate model. This is significant because removing the rescaling process results in a more transparent model and less reliance on UNHCR data.

\begin{figure}[t!]
       \centering \hspace{5mm}
       \includegraphics[width=0.5\textwidth]{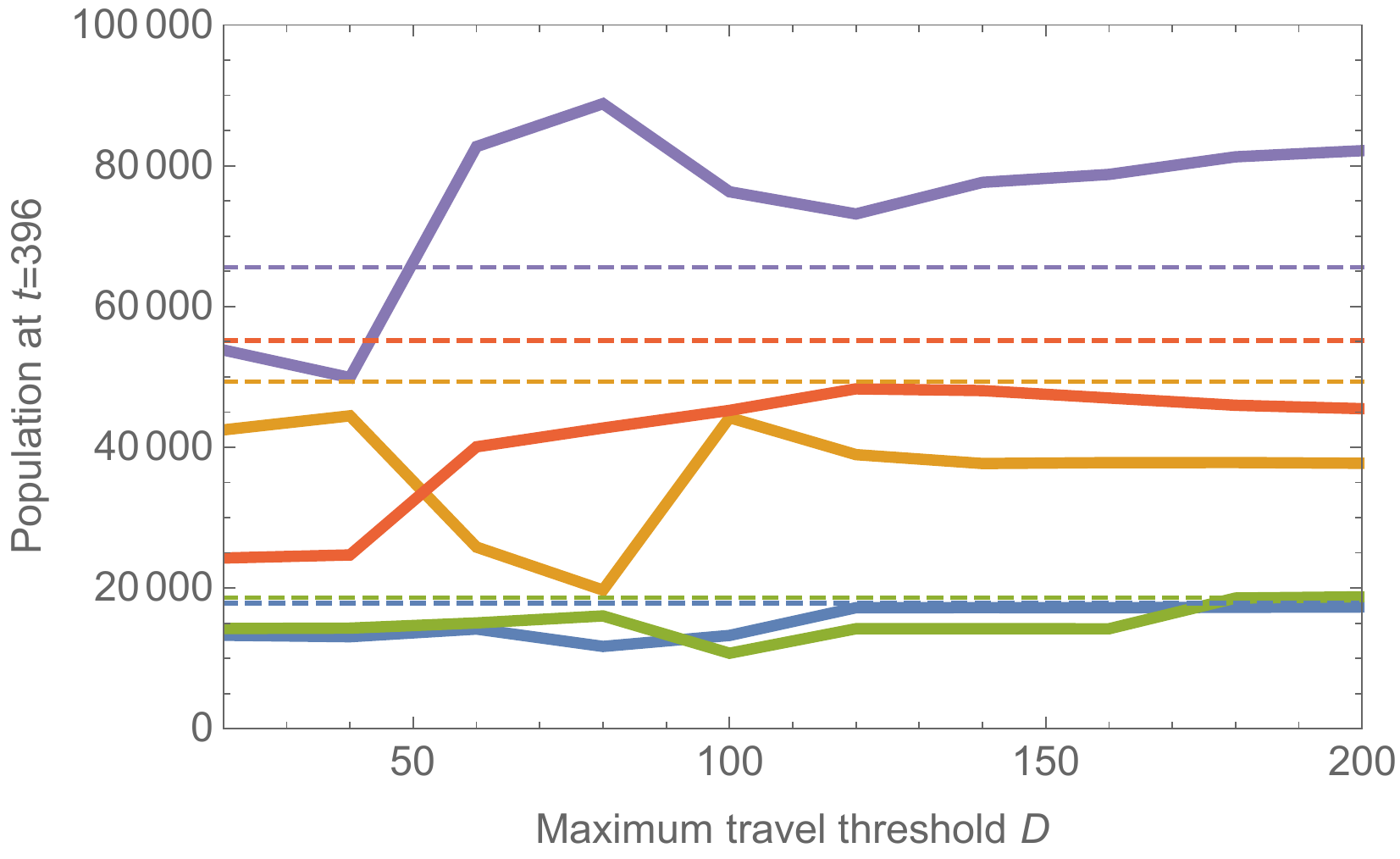}
       \includegraphics[width=0.4\textwidth]{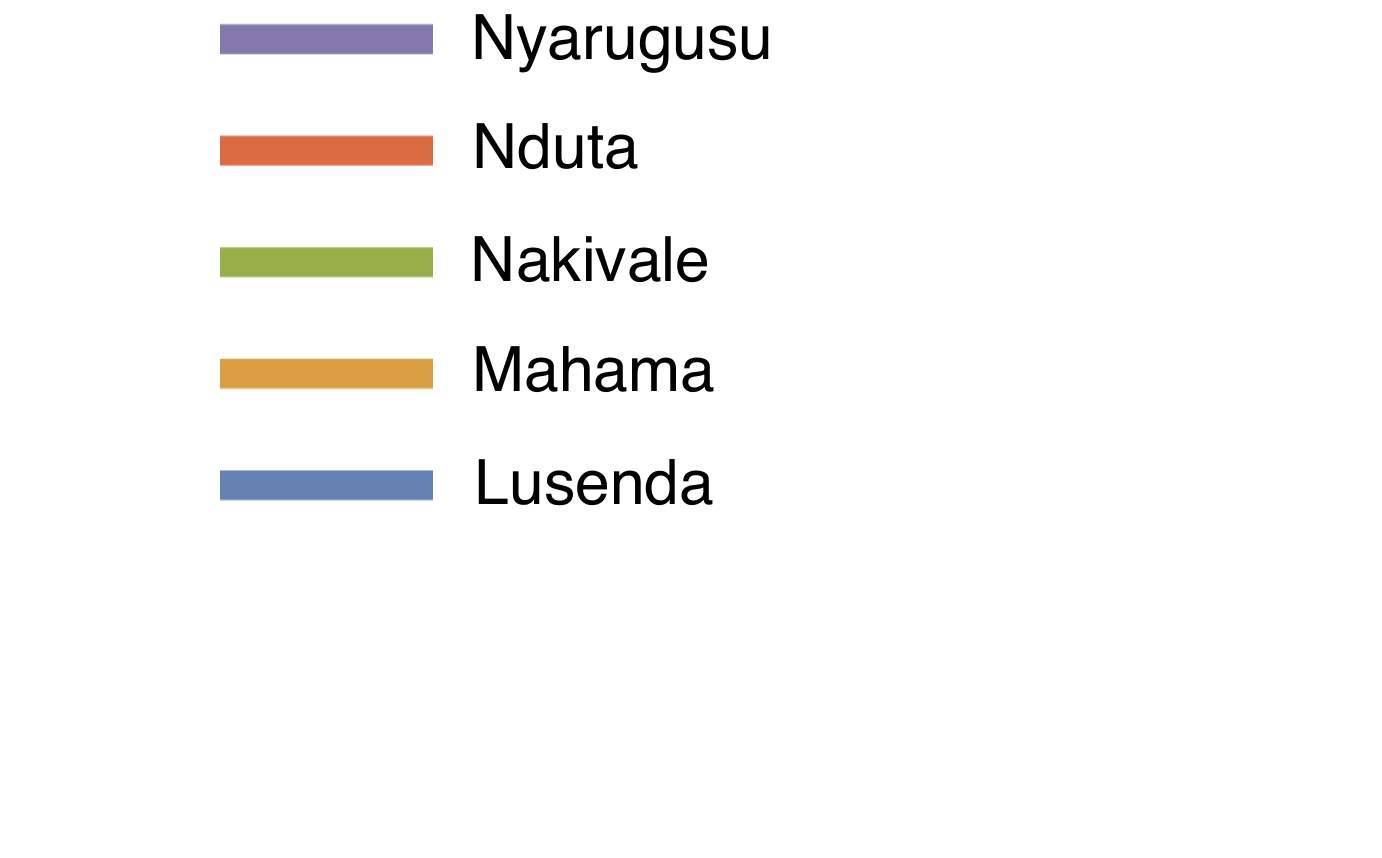}
              \caption{The  population of each camp at the final entry in the data set $(t=396)$ predicted by the Markov Chain: Camp Adjusted model as the maximum distance threshold $D$ is varied. The prediction for each camp is indicated by a different solid curve, the dashed lines indicate the population reported in the UNHCR data as given in \cite{Suleimenova}. Notably, as $D$ is varied the number of edges in the graph varies as described in Section \ref{sec2}.  } 
       \label{Fig5}
\end{figure}

A key parameter in the Markov chain models is the distance threshold $D$ which determines the maximum distance a refugee can travel in one day. This parameter is also a source of some uncertainty since, whilst our assumed choice $D=120$ (in units of km)  seems quite reasonable, it could conceivably vary by moderate margins depending on road quality, climate, and transportation options. Moreover, for the Camp Adjusted model as $D$ is varied the number of edges in the graph varies as described in Section \ref{sec2}. Thus it is prudent to explore how varying $D$ impacts the late time distribution of refugees. We show the impact of varying $D$ on the camp populations at $t=396$ (the final entry in the data set) in Figure \ref{Fig5}, the solid curves give the prediction for the {\em Markov Chain: Camp Adjusted} model, and the dashed lines indicate the value reported in the UNHCR data. Observe that for $D\ll 100$ then the fit to the late time data becomes marginally worse, however, for $D>100$ the results are relatively insensitive to changes in $D$ which provides some confidence that we need not be concerned by the exact value or indeed the possibility local variations in the maximum travel distance.

\section{Discussion: Comparison to Agent-Based Models}
\label{sec6}

We next discuss the differences between our Markov chain model and previous approaches using agent-based models, with specific reference to one state-of-the-art example due to Suleimenova, Bell, \& Groen \cite{Suleimenova} called {\em Flee}.
We will briefly summarize the main components of {\em Flee} and then compare our results to those obtained from running Flee (which is publicly available). 

{\em Flee}  \cite{Suleimenova}  implements a drastically different approach known as agent-based modelling \cite{Macal}. An agent-based model consists of a series of fixed states and a group of agents. Individual agents transition through these states according to the rules of the system, updating their current state at each integer timestamp. 
In modelling migration, a common approach is to define a weighted graph to contain the geographic information of a region of interest, and introducing one agent to represent each individual  the region. These agents then migrate between the vertices of the graph according to some predefined algorithm. This approach has been well-studied and utilised in areas such as disaster-driven migration \cite{Entwisle}, the ongoing conflict in Syria \cite{Latek}, and in \cite{Suleimenova}  applied to conflicts in Burundi, Mali, and the Central African Republic. 

\newpage

As in our  Markov chain model, {\em Flee}  \cite{Suleimenova}  consists of a graph storing geographic information and a prespecified algorithm used by refugees to move between cities of the graph. We noted earlier that we derive the planar graph (Figure \ref{Fig1}b) in our Markov chain model from {\em Flee}. Meanwhile, their agent-based model differs significantly from our  Markov chain model, because {\em Flee} treats each refugee in the simulation as an entirely separate entity making independent choices. Whereas we model refugee movements in terms of general probabilities of transitioning between states, {\em Flee} iterates over all agents in the system during every timestamp and simulates for each individual agent its expected movement through various cities. 

We noted earlier that the heuristic we use for properly initializing the population vectors at any given timestamp is derived from Flee's heuristic for generating new agents at each day of the simulation, the number of new agents added to the system is computed using real data about refugee numbers. This is acceptable because {\em Flee} aims to predict the end distribution of refugees over a network rather than the actual number of refugees displaced, as such, a rescaling of the distribution is performed at the end of every timestamp to ensure the total number of refugees in the camps in {\em Flee} matches the corresponding total according to UNHCR data, the same rescaling we use in three of the Markov chain models.

\begin{figure}[t!]
       \centering
      \vspace{-3mm}
       \includegraphics[width=0.45\textwidth]{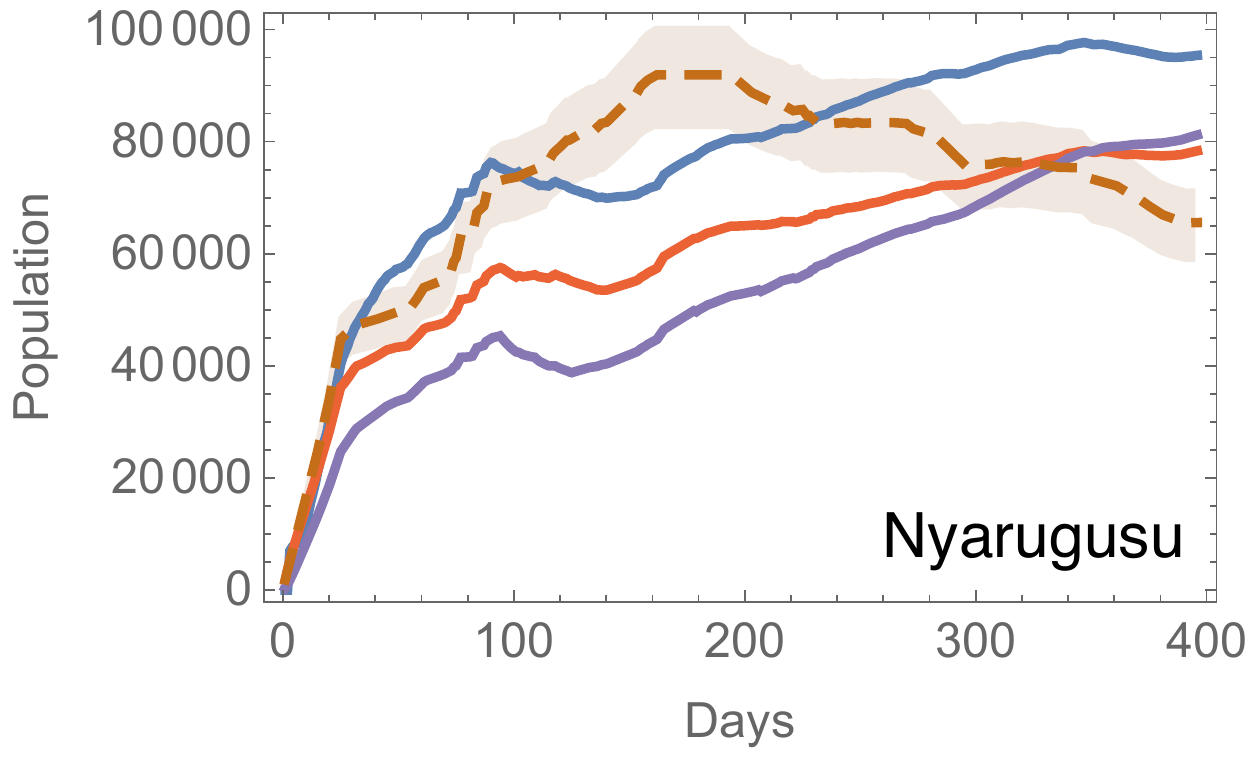}~
       \includegraphics[width=0.45\textwidth]{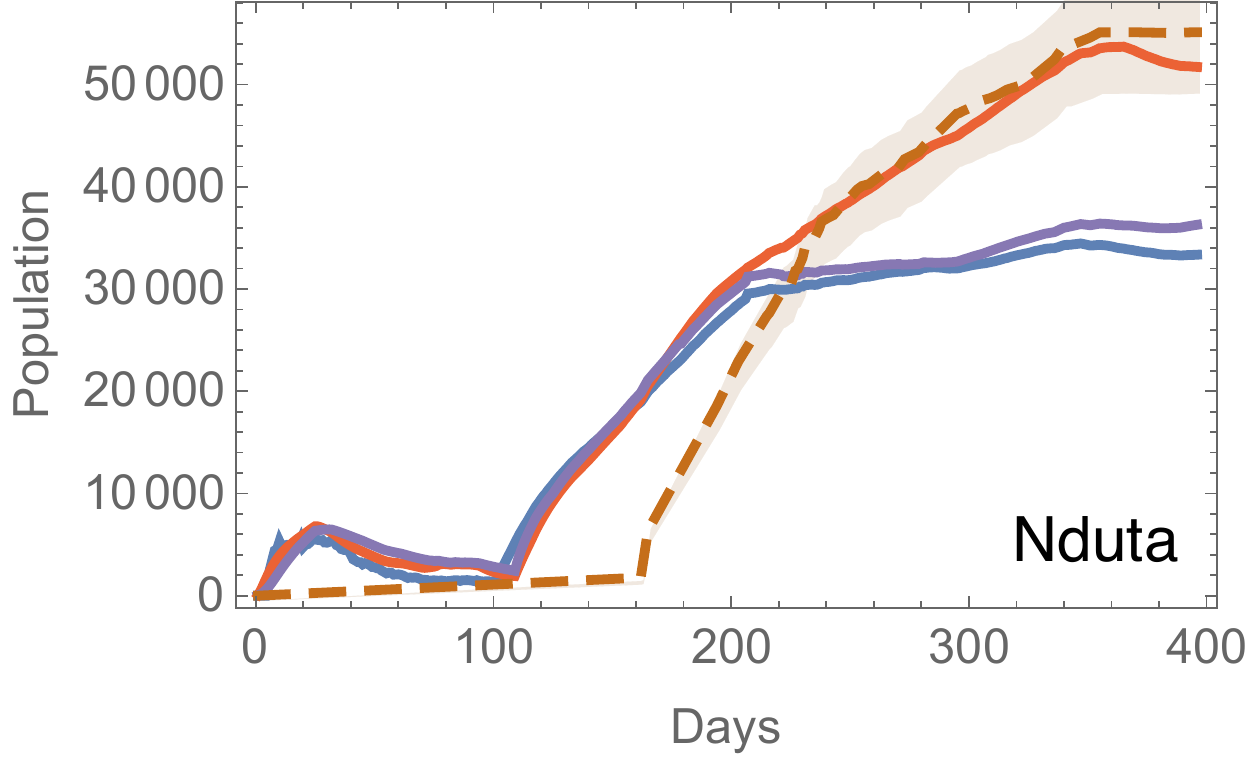}\\
       \includegraphics[width=0.45\textwidth]{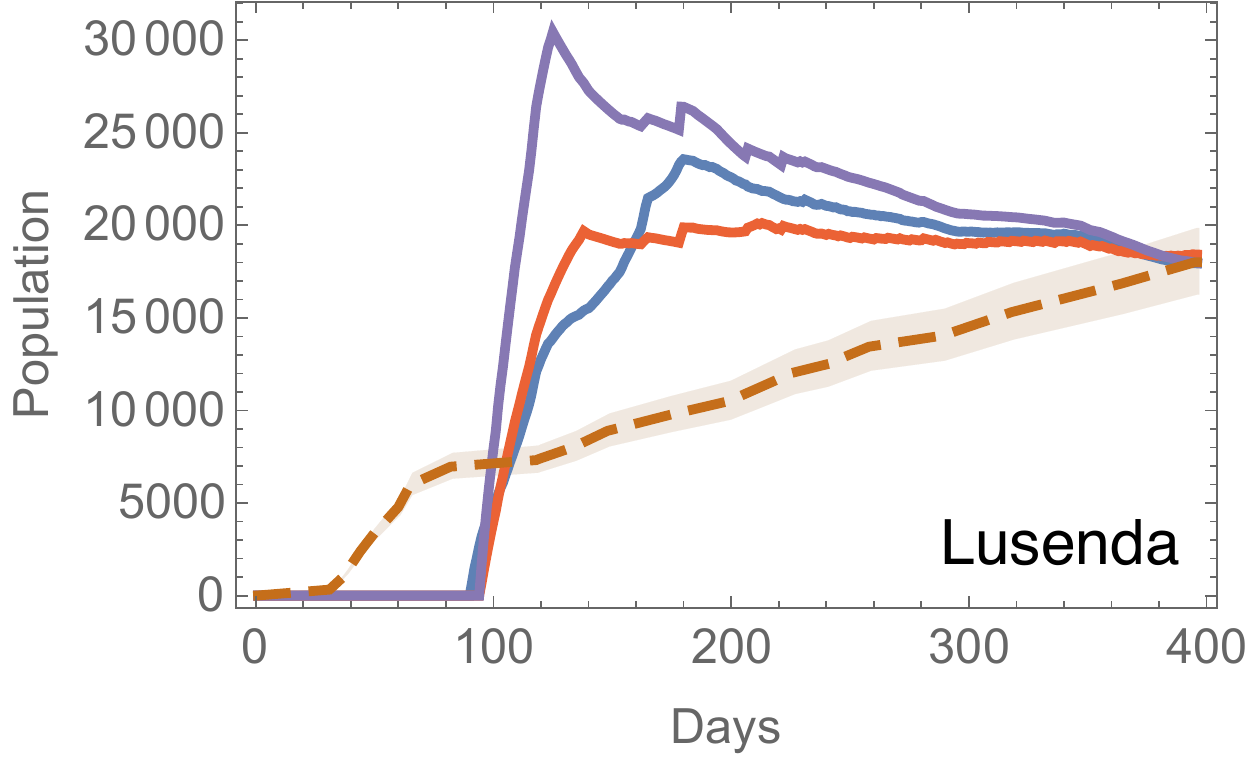}~
       \includegraphics[width=0.45\textwidth]{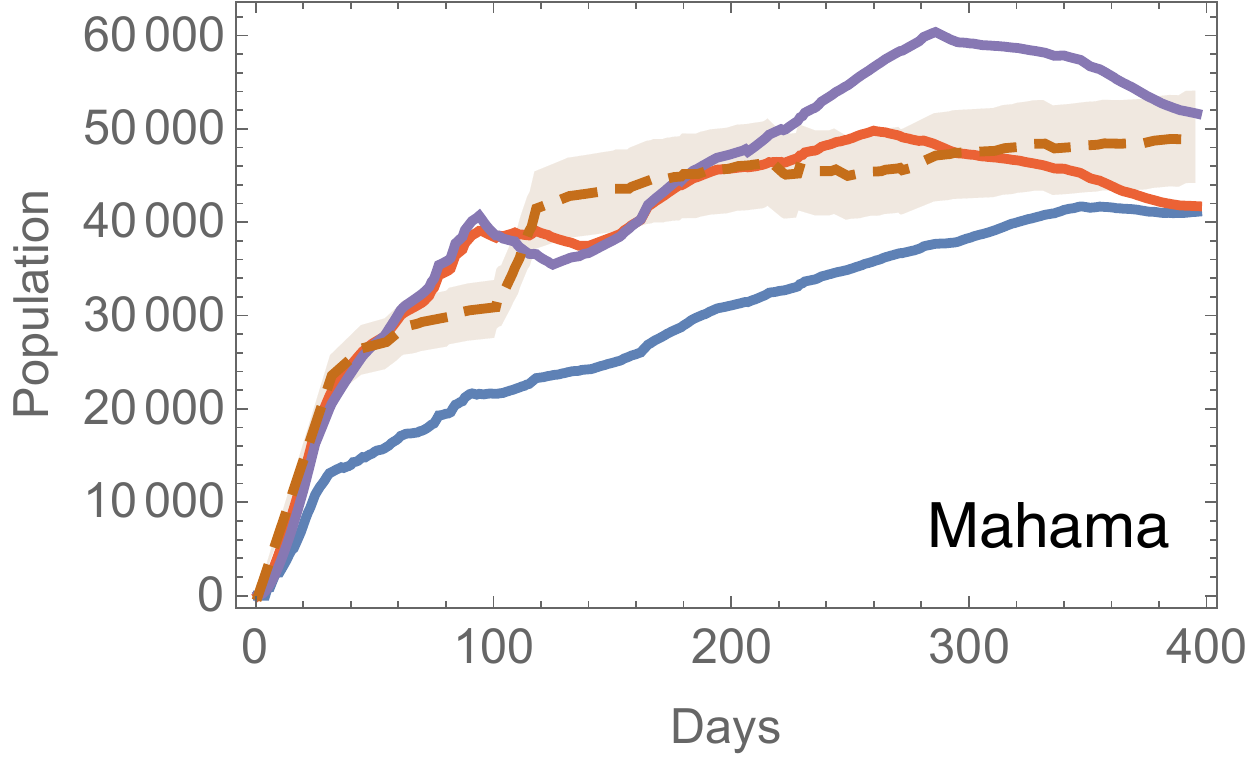}\\
       \includegraphics[width=0.45\textwidth]{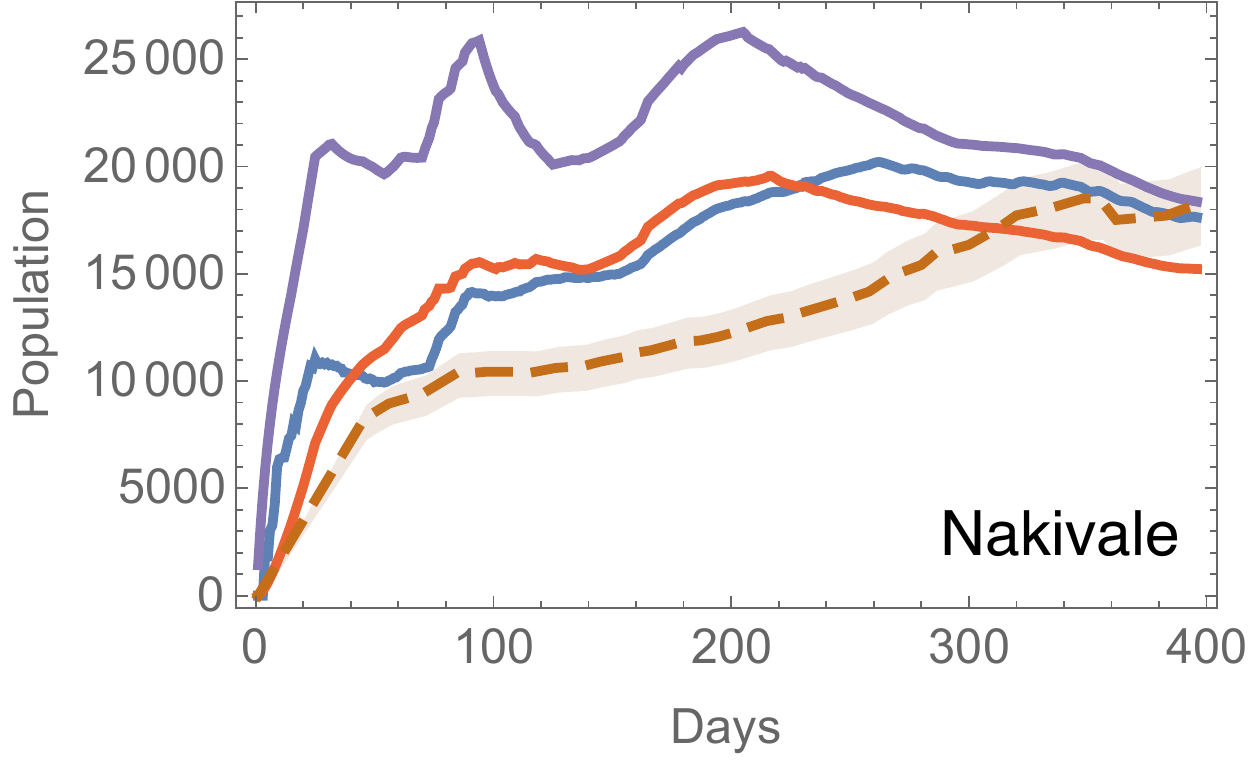}~
       \includegraphics[width=0.45\textwidth]{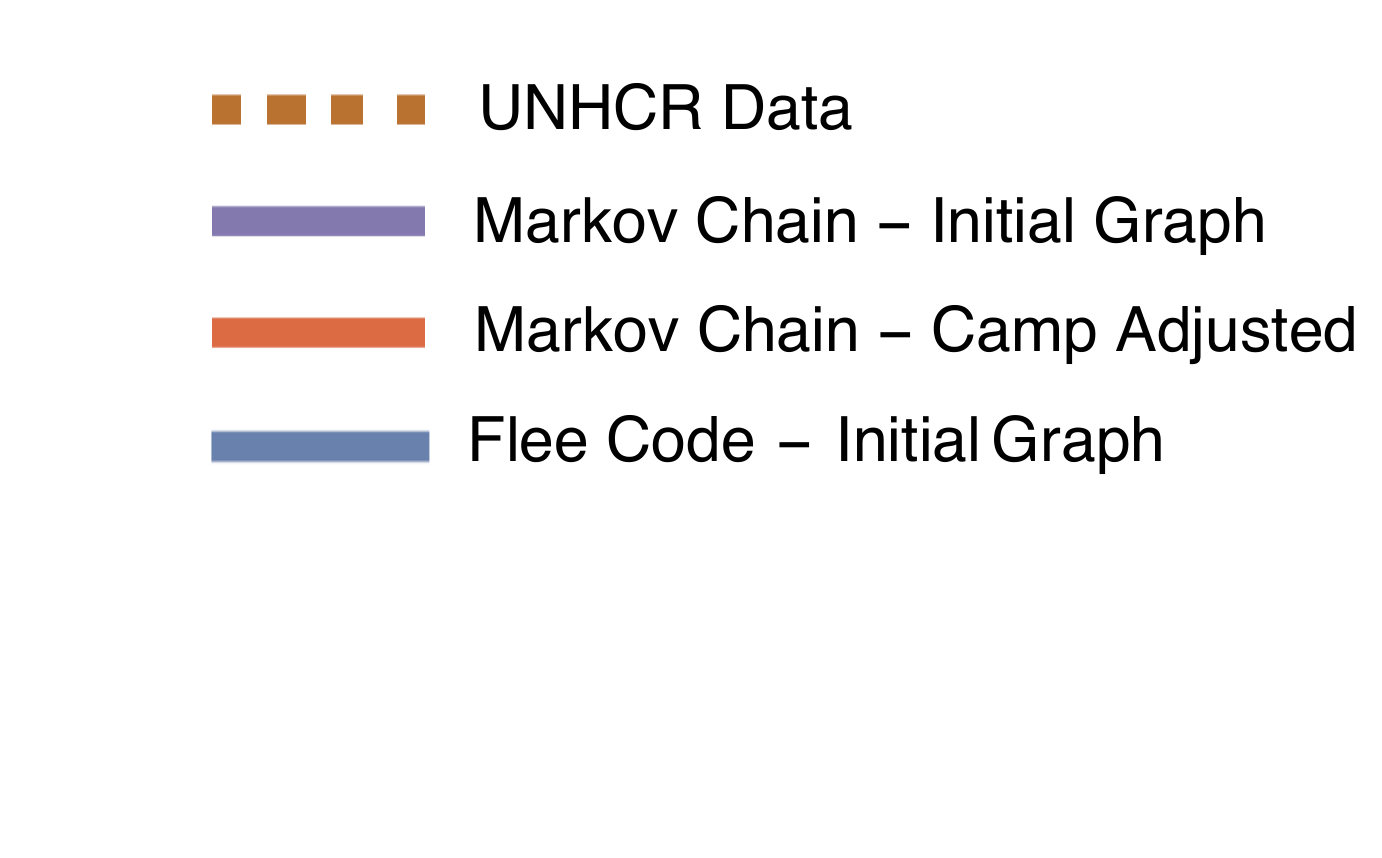}\\
      \vspace{-3mm}              \caption{Analogous to Figure \ref{Fig2} but here we compare our best model {\em Markov chain: Camp-adjusted} to the {\em Flee} model of SBG \cite{Suleimenova} and also to {\em Markov chain:  Initial Graph}. Note  {\em Flee} and {\em Markov chain:  Initial Graph} use the same graph, Figure \ref{Fig1}b,  whereas the  Camp-adjusted model uses the graph of Figure \ref{Fig1}c.       \label{Fig2b}} 
\end{figure}

Despite some similarities between {\em Flee} and our  Markov chain models, differences in algorithms lead to significantly different results.  We now compare the evolution of the camp populations and the ARD measures defined in eqns.~(\ref{ARD})-(\ref{395}) to quantitively compare our Markov chain models to Flee. Figure \ref{Fig2b} is analogous to Figure \ref{Fig2}, but illustrates instead the change in the populations of each camp as predicted by Flee (which match results presented in  \cite{Suleimenova}) to our Markov chain models implemented on the Initial Graph (using same graph as  \cite{Suleimenova}) and the {\em camp adjusted} graph (Figure \ref{Fig1}c). 

\begin{figure}[t!]
       \centering
       \includegraphics[width=0.44\textwidth]{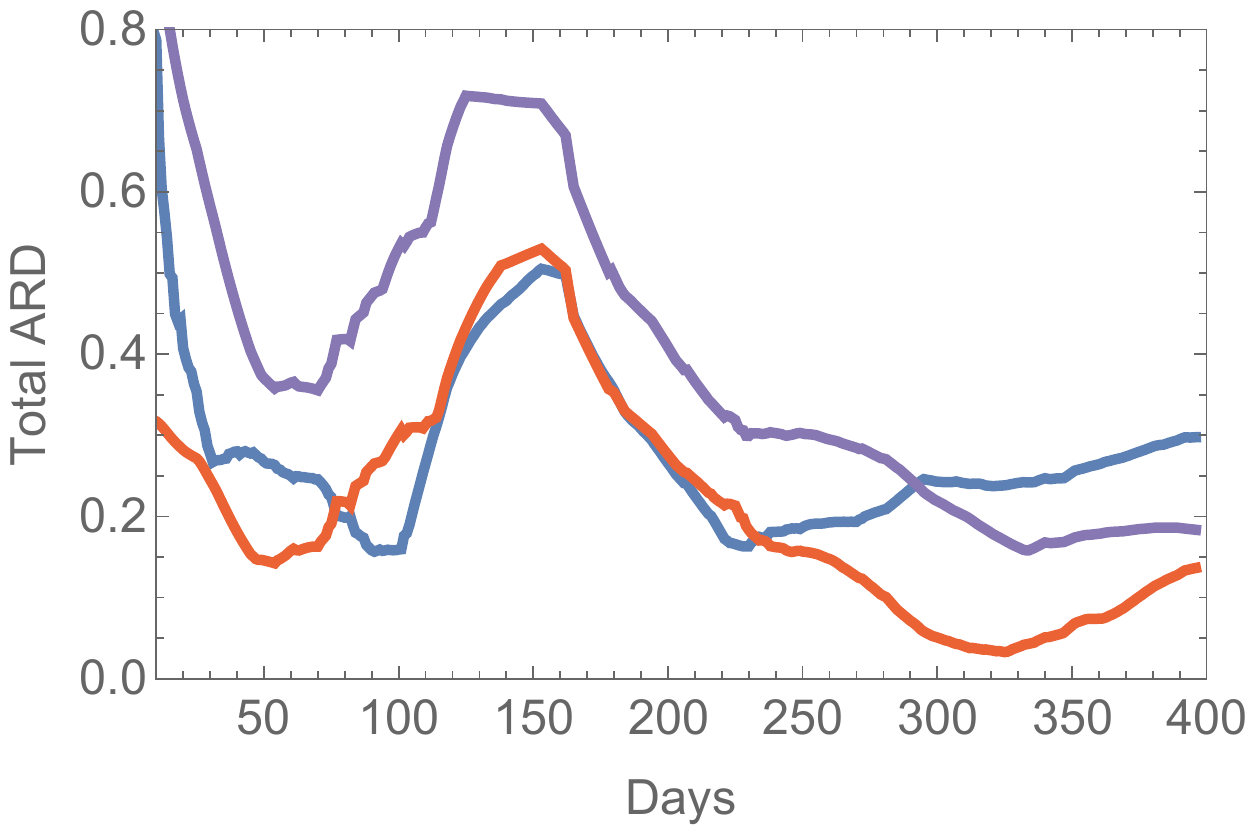}
              \caption{Comparison of the total  Averaged Relative Difference (ARD) on each timestamp (day) for the agent-based model `Flee' \cite{Suleimenova} and our Markov chain model applied to the  graph of Figure \ref{Fig2}b, alongside the Markov chain model applied to the camp-adjusted graph as in Figure~\ref{Fig2}c.      \label{Fig4}}  
 
\end{figure}

Additionally, in Figure \ref{Fig4} we show the total ARD as a function of timestamp for the three models displayed in   Figure \ref{Fig2b}, which gives provides a quantitive assessment of the relative goodness of fit of each model. It is noteworthy that by the end of the simulation, each of the five models has more or less converged to a constant value, and {\em Flee} has the highest total ARD.  The trends are captured concisely in Table \ref{tab2} which shows the all time Average ARD $\langle E\rangle$, and the Clean Average ARDs $\langle E\rangle_{30}$  and $\langle E\rangle_{100}$ for each model.

\vspace*{3mm}
\begin{table}[H]
\begin{tabular}{|c|c|c|c|c|c|}
\hline
        & Initial Graph & Camp-Adjusted & {\em Flee}      \\ \hline
$\langle E \rangle$   & 0.41 & 0.22      & 0.29        \\ \hline
$\langle E \rangle_{30}$ & 0.38 & 0.21      & 0.27        \\ \hline
$\langle E \rangle_{100}$& 0.37 & 0.21      & 0.28       \\ \hline
\end{tabular}
\caption{All time Average ARD $\langle E\rangle$, and Clean Average ARDs $\langle E\rangle_{30}$  and $\langle E\rangle_{100}$ for each model.}
\label{tab2}
\end{table}

The Camp-Adjusted Model and {\em Flee} display relatively stable average ARDs of approximately $0.21$ and $0.28$ over time, while the other models take longer to converge to lower average ARD values. In particular, note that in terms of long-term predictive power, which is the main focus of refugee models, the Markov chain: Camp Adjusted  model outperforms both Flee and the Markov Chain: Initial Graph. The Camp-Adjusted  model  has a value of  $\langle E \rangle_{100}$ for the is roughly $76\%$ that of Flee, implying a $24\%$ reduction in the long-term error.

      \vspace{-5mm}
\clearpage

In \cite{Suleimenova} to quantify their results the authors use instead the camp-wise ARD $E_i(t)$ for a given day as the main metric of success  in modelling the Burundi crisis. Here we have focused instead on the total ARD which we believe is a clearer indicator of overall modelling accuracy, however the camp-wise ARDs for each model are provided in the Appendix to provide a more direct comparison and one can draw similar conclusions on the goodness of fit of each model from looking at the ensemble of the individual camp ARDs. 

Thus we conclude that  Markov chain models can give a modest improvement on the goodness of fit compared to {\em Flee}, moreover, Markov chains yield a significantly more efficient simulation tool than an agent-based approach in terms of runtime and code length.  In terms of code length, the components necessary to make complex agent-based models such as {\em Flee} operate are spread over multiple folders, dozens of files, and thousands of lines in code \cite{Suleimenova}. In contradistinction, each of our  Markov chain models is contained within a single file and less than 500 lines of code. While program length is no measure of efficiency or suitability, the maintenance costs of code and the likelihood of bugs are directly correlated with program length. 

To quantify the complexity of the agent-based {\em Flee}, consider a total of $R$ refugees and $n$ cities in a region of interest, for a model which runs $t$ time steps. At each timestamp, {\em Flee} iterates through all agents in the system to update their status, because agent-based models treat each agent separately. Furthermore, each agent runs through the list of $n$ cities before deciding which city to migrate to, thus each of $R$ agents takes $O(n)$ time to update at each timestamp, leading to a total time complexity of at least $O(nRt)$. 

Analyzing the overall time complexity of our  Markov chain models is slightly harder. If there are $n$ cities in the region of interest, then we noted in Section \ref{sec2} that our  graph modification algorithms utilizing Floyd-Warshall run in $O(n^3)$ time, however, these algorithms only run once, at the very beginning of the simulation, to establish the graph. Meanwhile, we noted in Section \ref{sec3} that we compute the transition probabilities $P_{ji}$ in $O(n^3D)$ time. Once the transition probabilities are computed once, they do not need to be recomputed, as we reuse the same stochastic matrix $A(t)$ over multiple timestamps, unless a fundamental change is made to the cities in the graph, for instance, a neutral city turning into a conflict site changes the intermediate probabilities and therefore the transition probabilities of the system. The number of such updates to vertices is on the order of the number of vertices $n$, because neutral cities can become conflict sites, but the opposite is much rarer. As a result, computing transition probabilities takes around $O(n)\cdot O(n^3D)=O(n^4D)$ time. Finally, at each timestamp, the Markov chain only performs a simple multiplication of an $n\times n$ matrix with an $n$-dimensional vector, which only takes $O(n^2)$ operations, thus updates to the system comprise a total of $O(n^2t)$ time. Thus we estimate the time complexity of the Markov chain models is $O(n^3 + n^4 D + n^2t)$.

In the case of Burundi, $R\sim 10^5$ refugees, $t=396\sim10^2$ days, and there are $n\sim 10$ vertices, while $D\sim10^2$, thus {\em Flee} runs in $O(nRt) =O(10^8)$ operations, while the Markov chain models are $O(n^3 + n^4D + n^2t) = O(10^6)$ operations, because the $n^4D$ term dominates the asymptotic. Indeed, this order of magnitude difference in runtimes is apparent when executing these different simulations on a standard desktop or laptop.  In general, migration models will typically only involve significant urban centers, so the value of $n$ is reasonably small, while $R$ can be incredibly large, making agent-based model less feasible than Markov chains implementations.


\section{Concluding Remarks}
\label{sec7}
In this work we have implemented Markov chain models of refugee migration in refugee crises based on new heuristics about the connections between refugee movements and local geography. By applying our model to the Burundi refugee crisis and comparing with an existing agent-based model of that same crisis, we concluded that our  approach using Markov chains was more efficient, reduced unnecessary complications in agent-based modelling, and exhibited a $24\%$ reduction in long-term prediction errors. The agent-based model {\em Flee} was explicitly validated against three separate conflicts (Burundi, Mali, and the Central African Republic) in \cite{Suleimenova}, whereas for brevity we have restricted our analysis here to just the Burundi conflict. However, we highlight in passing that the Markov Chain models developed here similarly provide good fits (comparable to Flee) to these other two conflicts when applied to the data and graphs of Mali and the Central African Republic presented in \cite{Suleimenova}.

It should be noted that collecting data on refugee registrations at camps is a challenging task and it is highly feasible that there could be sizeable discrepancies between the UNHCR data and real camp populations. Moreover, refugees that leave a camp for a new destination, say another camp or perhaps a relatives house in a neutral city, are likely not to see deregistering as a priority and thus the data my not accurately reflect departures from a camp.  We have endeavored to account for these potential discrepancies in the data by including a 10\% error margin in Figures \ref{Fig2} \& \ref{Fig2b}, however this is a rather blunt heuristic and a better understanding of the intrinsic errors in the data would be highly desirable. Applying our model to higher quality and larger data sets would allow us to better refine the Markov chain models presented here.

As a result of the recent increases in refugee crises, there are many reasons as to why improved refugee models would be useful. Accurate refugee models would allow us to  predict the number of refugees who will arrive at a particular area or city, the date at which they will arrive, and the distribution of these refugees across multiple regions, a few days or even weeks in advance, given good information about the causes of the migration. This would be extremely beneficial as accurate modelling could allow refugee and governmental organizations to determine where to best distribute aid resources to maximize impact and efficiency. Predictive models would furthermore allow cities and regions to take the appropriate measures to accommodate a large influx of people seeking shelter, which can typically lead to highly disruptive situations. Indeed, with the onset of global climate change,  increasing inequality, and decreasing global stability it is highly likely that the number of displaced peoples will increase significantly in the future \cite{Myers}. Thus there is a growing need for better simulations and in this work we have advocated for adopting Markov chain based models in order to more accurately and efficiently model these crises.

\vspace{-3mm}
\section*{Acknowledgements}

We thank William Jones, Laura Schaposnik, and especially Derek Groen for invaluable discussions, as well as Tanya Khovanova and Claude Eicher for comments on the draft.  JU is grateful to the Simons Center for Geometry and Physics (Program: Geometry and Physics of Hitchin Systems) and New College, Oxford for their hospitality and support. This work was completed as part of the MIT PRIMES Program.


\section*{Appendix: ARDs for Individual Refugee Camps}

\begin{figure}[b!]
       \centering
       \includegraphics[width=0.47\textwidth]{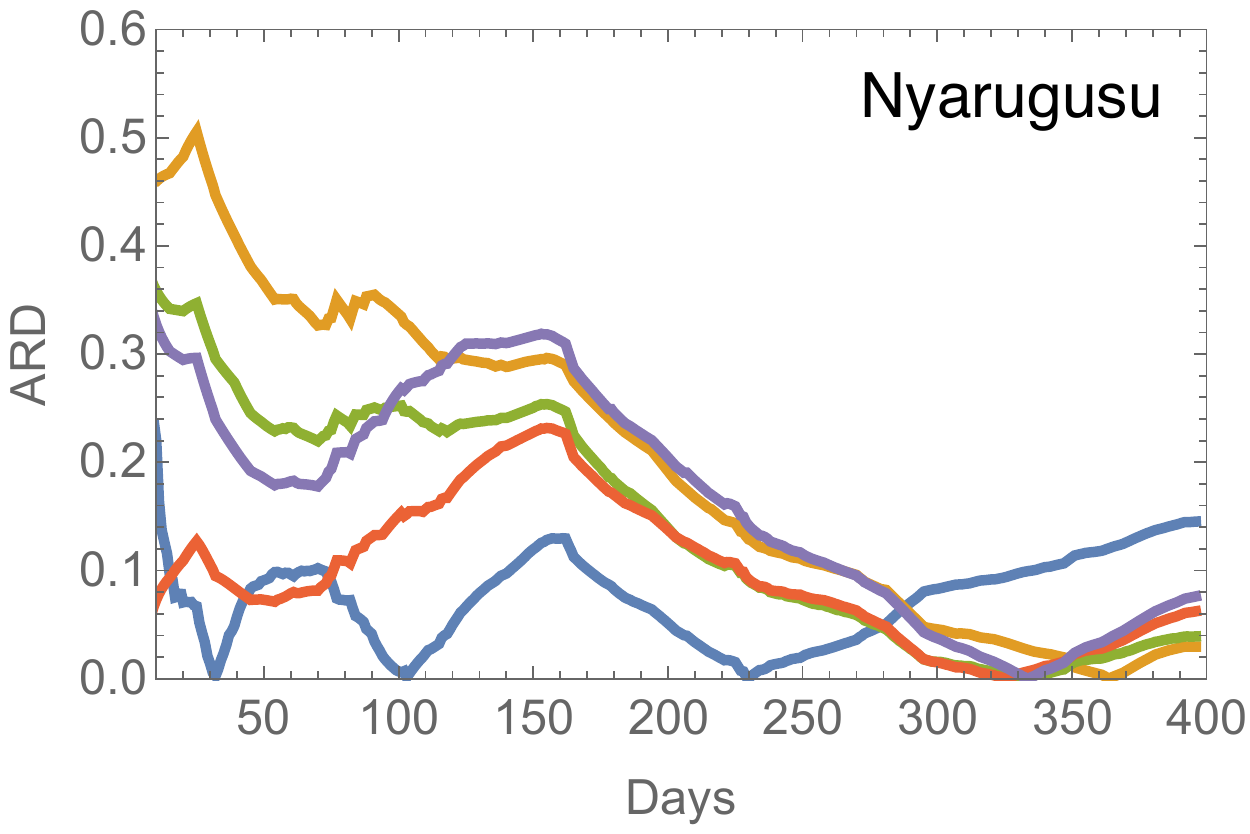}~
       \includegraphics[width=0.47\textwidth]{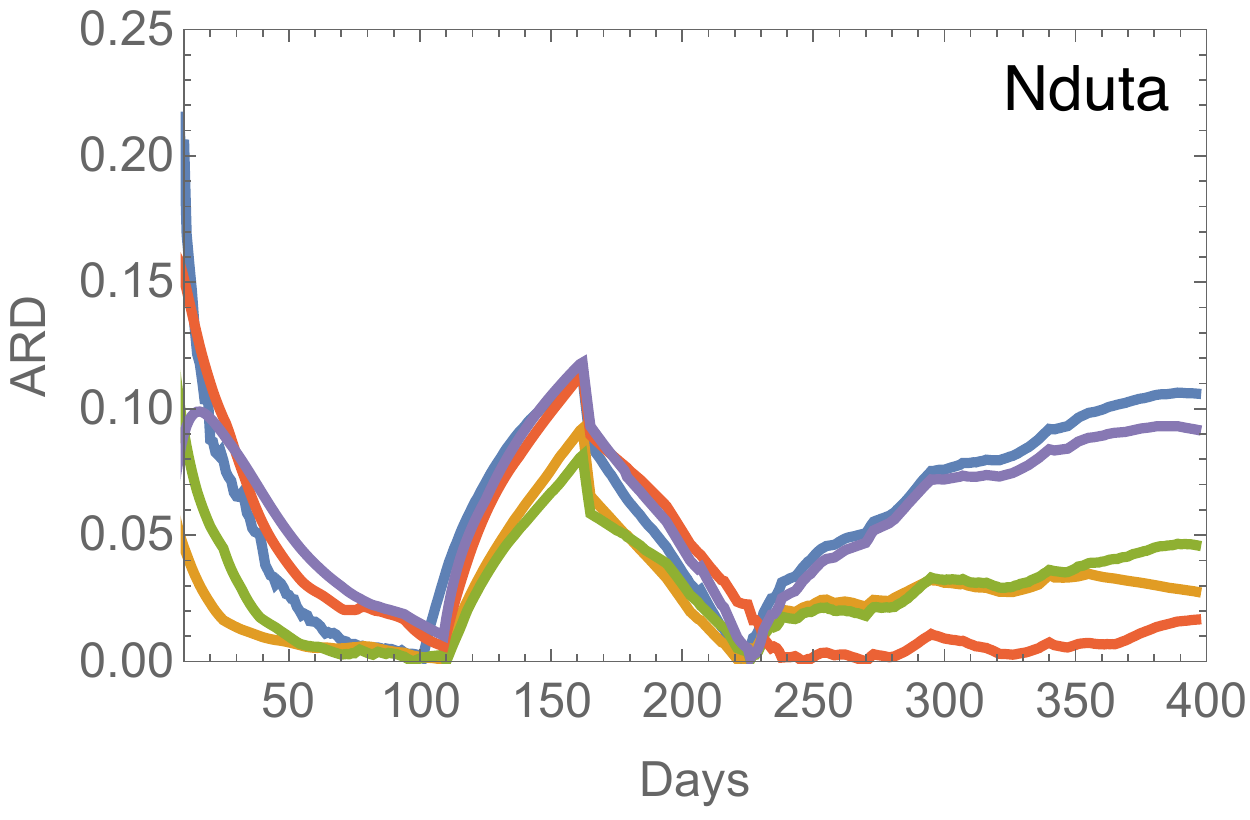}\\
       \includegraphics[width=0.47\textwidth]{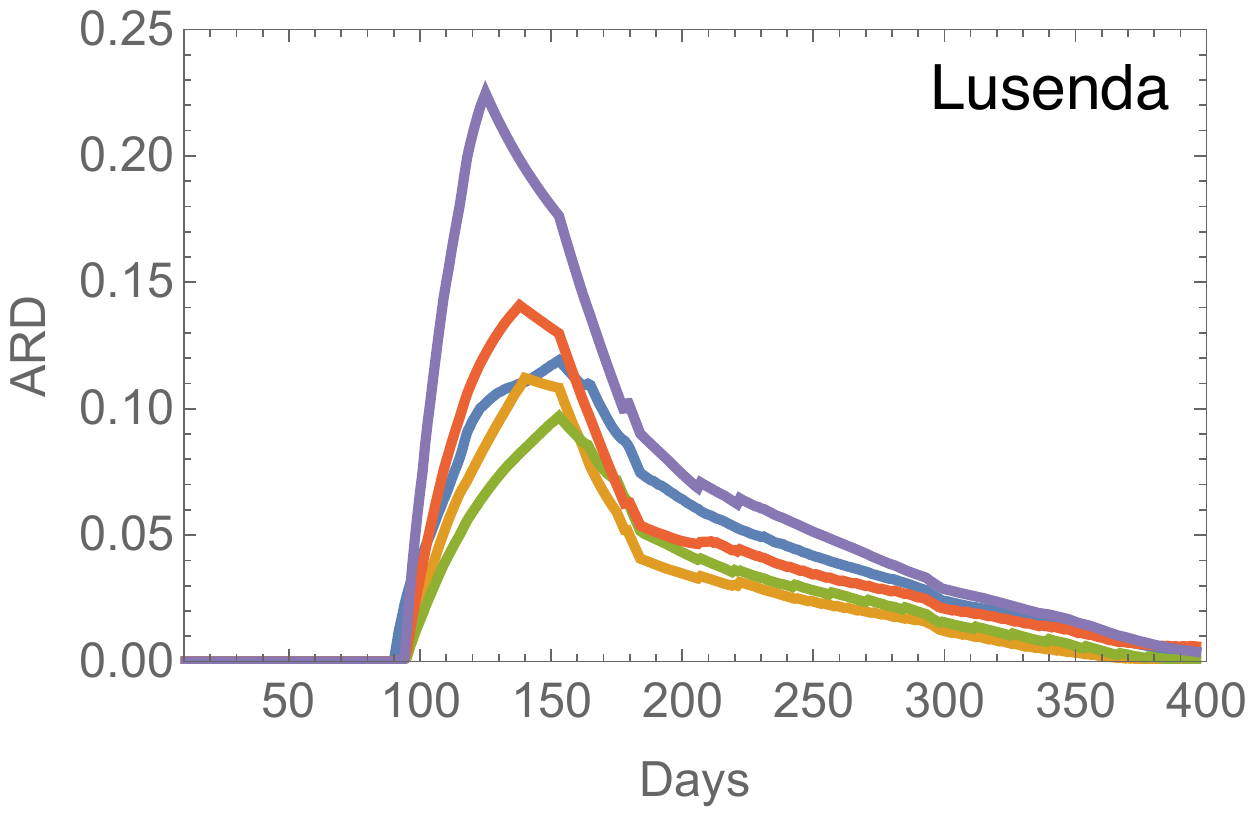}~
       \includegraphics[width=0.47\textwidth]{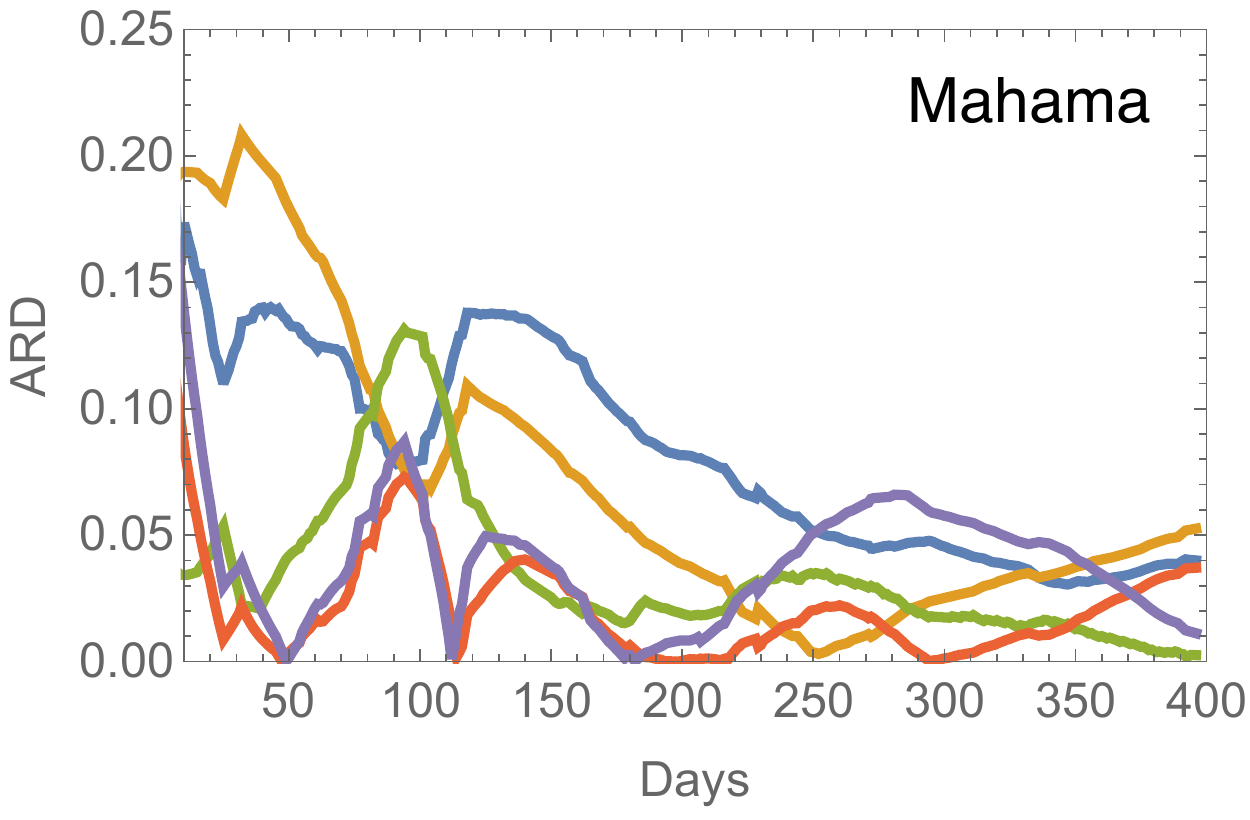}\\
       \includegraphics[width=0.47\textwidth]{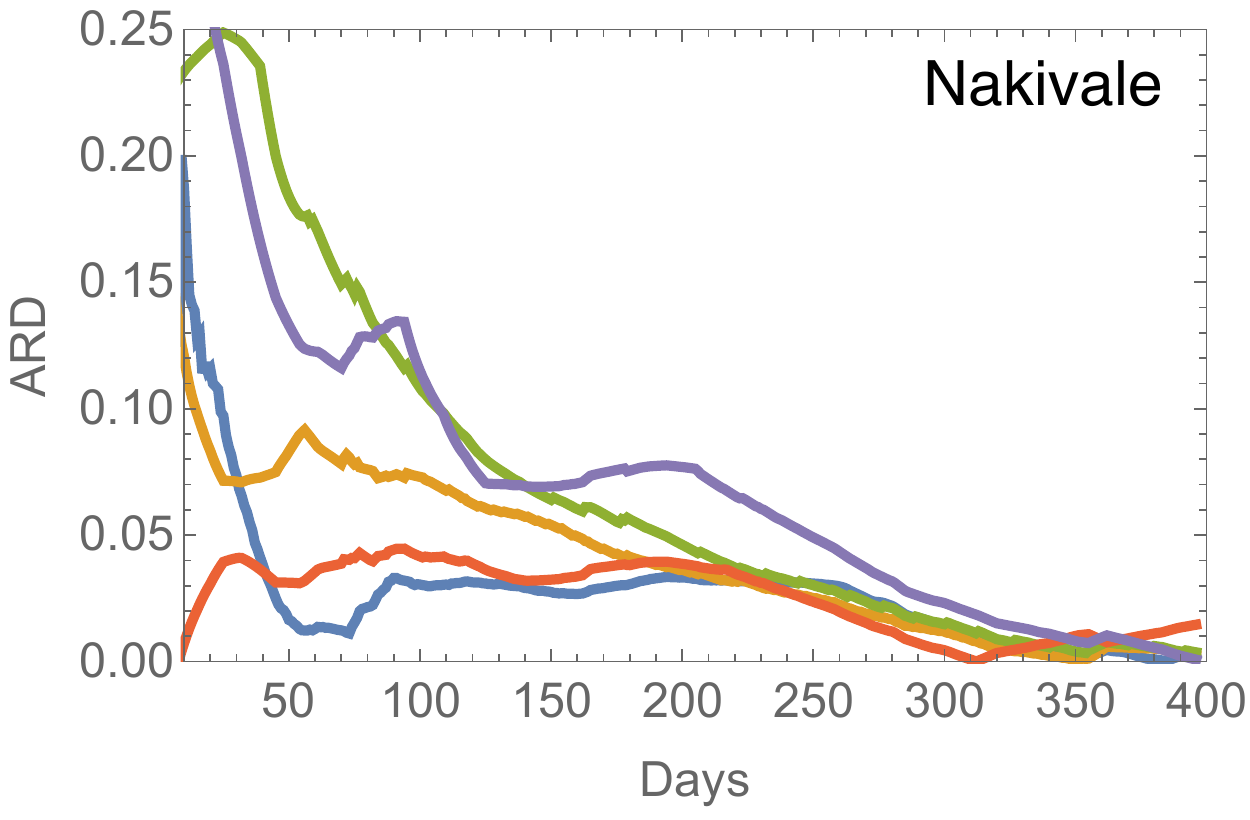}~
       \includegraphics[width=0.47\textwidth]{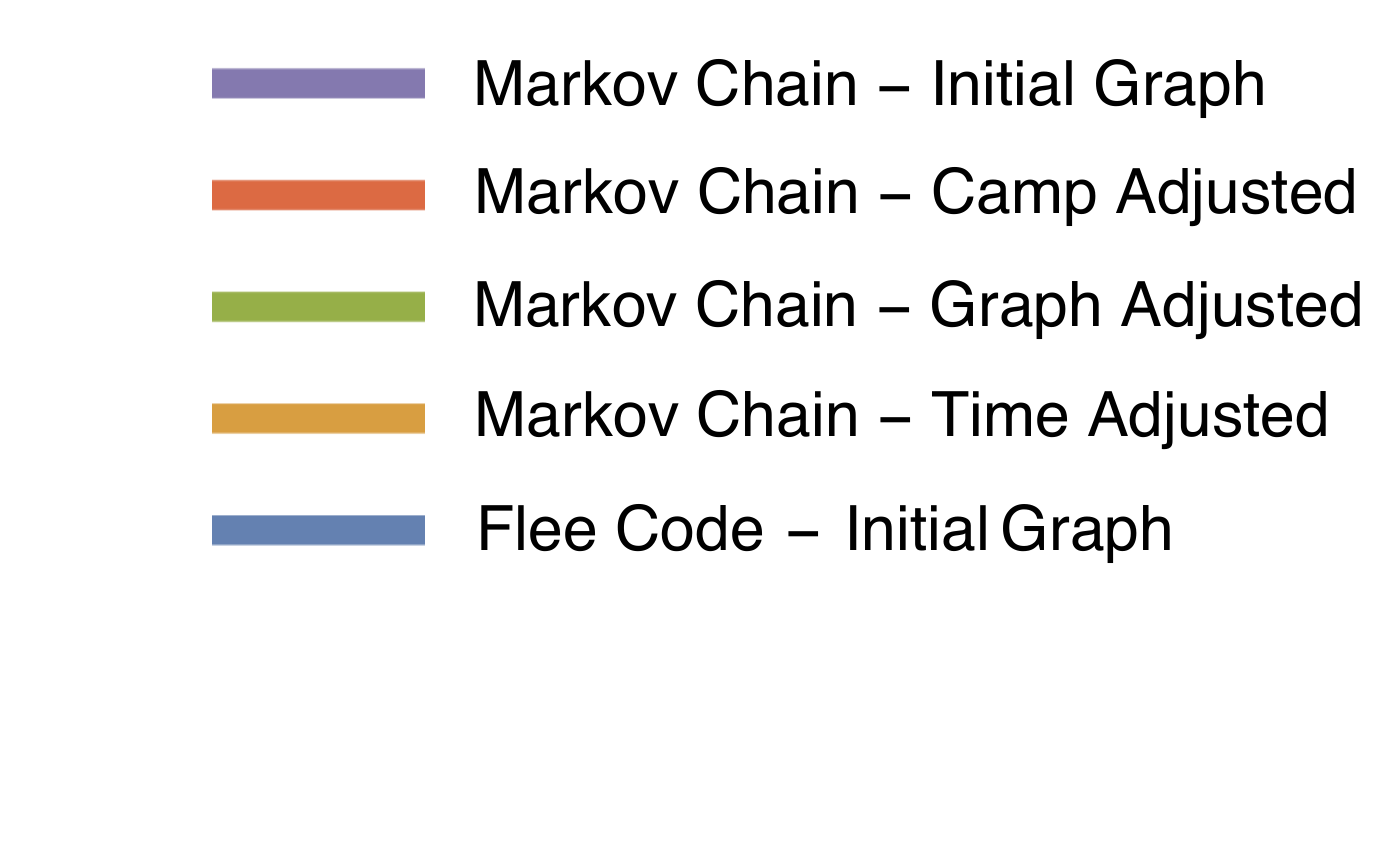}
              \caption{Averaged Relative Difference (ARD) for each model for each Camp.} 
       \label{FigA}
\end{figure}

In this appendix we present the ARD for each camp as a function of timestamp. The ARD for a camp gives a good measure for the amount of error present in modelling a single camp as a fraction of total refugee population.  Note that Figures \ref{Fig3} \&  \ref{Fig4}, which show the total ARD for each model on each day essentially ``sum'' of the five graphs in Figure \ref{FigA}.

Figure \ref{FigA} shows the ARDs in each of the five refugee camps of each of the five models, the  four Markov chain models developed here, and SBG's {\em Flee} model \cite{Suleimenova}. Observe that in two of the camps, Lusenda and Nakivale, all five models perform quite well, leading to an almost indistinguishable long-term ARD of approximately zero. However, this trend does not hold in the other three camps.

In Nduta, all three models utilizing our  graph modification algorithm from Section \ref{sec2} perform significantly better than than {\em Flee} and the Markov Chain Model: Initial graph, which both use the unmodified graph. In Nyarugusu, {\em Flee} initially produces relatively accurate results, but the four Markov chain models eventually converge to a more accurate prediction. Finally, in Mahama, it is not immediately clear which models yield more accurate results, although the Graph-Adjusted and Time-Adjusted  Models more consistently produce smaller ARDs. 

This camp specific ARD is the main tool used in \cite{Suleimenova} for quantifying the success of their model of the Burundi crisis and thus \ref{FigA} allows for a direct comparison with the results of \cite{Suleimenova}. Moreover, an informal inspection of the ensemble of camp ARD graphs indicates similar conclusions as found in the main text, specifically that the Markov chain models  provide a modest improvement in the fit to the data compared to Flee and that the Markov Chain: Camp Adjusted model seems the best fit out of all of the models studied. In the main text these conclusions were quantitively supported by an analysis of the total ARD and Averaged ARGs, see Figures \ref{Fig3} \&  \ref{Fig4} and Tables \ref{tab1} \& \ref{tab2}.



\begin{thebibliography}{}

\bibitem{UNHCR}
United Nations High Commissioner for Refugees [n.d.]~{\em Figures At a Glance.} Retrieved January 24, 2018, from  http://www.unhcr.org/en-us/figures-at-a-glance.html


\bibitem{Ravenstein}
E.~G.~Ravenstein,  {\em The Laws of Migration,}  Journal of the Statistical Society of London. Vol. 48, No. 2 (Jun., 1885), pp.~167-235.

\bibitem{Lee}
E.~Lee, {\em  A theory of migration}, Demography 3, 471-486 (1966).


\bibitem{Willekens}
F.~Willekens, {\em Migration flows: Measurement, analysis and modeling}, International Handbook of Migration and Population Distribution, Springer Netherlands (2016) pp.~225-241.


\bibitem{Groen}
D.~Groen, {\em Simulating Refugee Movements: Where Would You Go?}  Procedia Computer Science, Volume 80, 2016, Pages 2251-2255. 

\bibitem{Gulden}
T.~Gulden, J.~Harrison, A.~Crooks, {\em Modeling Cities and Displacement through an Agent-based Spatial Interaction Model}, Computational Social Sciences (2011). 

\bibitem{Klabunde}
A.~Klabunde, and F.~Willekens, {\em Decision-Making in Agent-Based Models of Migration: State of the Art and Challenges
}, Eur.~J.~Population (2016) 32: 73. 


\bibitem{Suleimenova}
D.~Suleimenova, D.~Bell, and D.~Groen. {\em A generalized simulation development approach for predicting refugee destinations,}  Scientific reports 7, no.~1 (2017): 13377.


\bibitem{Christou}
C.~Christou, {\em Simulation of Human Migration Based on Swarm Theory}, 
2010 13th International Conference on Information Fusion, Edinburgh, 2010, pp.~1-8.


\bibitem{Kniveton}
D.~Kniveton, C.~Smith and S.~Wood, {\em Agent-based model simulations of future changes in migration flows for Burkina Faso}, Global
Environmental Change 21, 34-40 (2011).

\bibitem{Johnson}
R.~Johnson, T.~Lampe, and S,~Seichter, {\em Calibration of an agent-based simulation model depicting a refugee camp scenario}, 
Proceedings of the 2009 Winter Simulation Conference (WSC), 1778-1786 (2009).

\bibitem{Sokolowski}
J.~Sokolowski, C.~Banks and R.~Hayes, {\em  Modeling population displacement in the Syrian city of Aleppo}, Proceedings of the
2014 Winter Simulation Conference, 252-263 (2014).


\bibitem{Macal}
See for instance,
C.~Macal and M.~North, {\em Tutorial on agent-based modeling and simulation}, Simulation conference: 2005 proceedings of the winter IEEE (2005).



\bibitem{Blumen}
I.~Blumen, M.~Kogan, and P.~McCarthy, {\em The Industrial Mobility of Labor as a Probability Process}, Ithaca: Cornell University Press (1955).

\bibitem{Salkin}
M.~Salkin, T.~Lianos and ~Q.~Paris, {\em Population Predictions for the Western United States: A Markov Chain Approach}, Journal of Regional Science, 15: 53-60 (1975).

\bibitem{Goodman}
L.~Goodman,  {\em Statistical Methods for the Mover-Stayer Model,}  Journal of the American Statistical Association, vol.~56, no.~296, 1961, pp.~841-868.

\bibitem{Brown}
L.~Brown,  {\em On the Use of Markov Chains in Movement Research,}  Economic Geography, vol.~46, 1970, pp.~393-403. 

\bibitem{Hirst}
M.~Hirst,  {\em A Markovian Analysis of Inter-Regional Migration in Uganda,}  Geografiska Annaler. Series B, Human Geography, vol.~58, no.~2, 1976, pp.~79-94.

\bibitem{Nagurney}
A.~Nagurney, {\em Migration equilibrium and variational inequalities}, Economics Letters, vol.~31, no.~1, (1989) 109-112.


\bibitem{Pan}
J.~Pan, and A.~Nagurney
{\em Using Markov chains to model human migration in a network equilibrium framework,} 
Mathematical and Computer Modelling,
vol.~19, no.~11 (1994)  31.

\bibitem{kim}
H.~Kim and H.~Y.~Song,
{\em Formulating Human Mobility Model in a Form of Continuous Time Markov Chain} 
Procedia Computer Science, vol.~10 (2012) 389-396.



\bibitem{Gagniuc}
P.~Gagniuc, {\em Markov Chains: From Theory to Implementation and Experimentation,}   John Wiley \& Sons. (2017) pp.~9-11. ISBN 978-1-119-38755-8.

\bibitem{Isaacson}
See for instance,
D.~Isaacson and R.~Madsen,  {\em Markov Chains Theory and Applications},  Wiley, New York (1976).


\bibitem{Baumann}
H.~Baumann, and W.~Sandmann, {\em Structured Modeling and Analysis of Stochastic Epidemics with Immigration and Demographic Effects}, PLoS ONE 11(3): e0152144 (2016).


\bibitem{internet}
L.~Muscariello, M.~Mellia,~M.~Meo, M.~Ajmone Marsan, and R.~Lo Cigno
{\em Markov models of internet traffic and a new hierarchical MMPP model},
Computer Communications, vol,~28, no.~16, (2005),  1835-1851.

\bibitem{PageRank}
L.~Page, {\em Method for node ranking in a linked database}, U.S.~Patent 6285999.

\bibitem{finance}
See for example, M.~Kijima, {\em Stochastic Processes with Applications to Finance}, New York: Chapman and Hall/CRC (2002) ISBN:9781420057607. 

\bibitem{Djordjevic}
I.~Djordjevic {\em Markov Chain-Like Quantum Biological Modeling of Mutations, Aging, and Evolution}, Life (2015) 5(3):1518-38. 

\bibitem{Bellman}
R.~Bellman, {\em The theory of dynamic programming}, 
Bull.~Amer.~Math.~Soc., 60 (6): 503-516 (1954).

\bibitem{FW1}
 R.~Floyd,  {\em Algorithm 97: Shortest Path}, Commun.~ACM. 5 (6): 345 (1962).

\bibitem{FW2}
 B.~Roy,  {\em Transitivite et connexite}, C.~R.~Acad.~Sci.~Paris.~249: 216-218 (1959).

\bibitem{FW3}
S.~Warshall,  {\em A theorem on Boolean matrices}, Journal of the ACM. 9 (1): 11-12 (1962).

\bibitem{google} Google (n.d.).~[Burundi]. Retrieved January 24, 2018, from https://www.google.com/maps/place/Burundi

\bibitem{Entwisle}
B.~Entwisle,  {\em et  al,}  {\em Climate  shocks  and  migration:  An  agent-based  modeling  approach},  Population  and  Environment  1-€"25  (2016).

\bibitem{Latek}
M.~Aatek,  S.~Rizi,  and A.~Geller,  {\em Verification  through  calibration:  An  approach  and  a  case  study  of  a  model  of  conflict  in  Syria},  Proceedings  of  the  2013  Winter  Simulation  Conference:  Simulation:  Making  Decisions  in  a  Complex  World,  1649-1660  (2013).

\bibitem{Myers}
N.~Myers, {\em Environmental refugees: a growing phenomenon of the 21st century}, Philos.~Trans.~Royal Soc.~B 357.1420 (2002): 609-613.





	\end{thebibliography}
\end{document}